\documentclass[review,authoryear,12pt]{elsarticle}

\usepackage{multirow}
\usepackage{url}
\usepackage{float}
\usepackage[section]{placeins}
\usepackage{amssymb}
\usepackage{amsmath}
\usepackage{lineno}
\usepackage{lipsum}
\usepackage{algorithmic}
\usepackage{mathtools}
\usepackage{subfig}
\usepackage{changepage}
\usepackage{xcolor,graphicx}
\usepackage[norelsize,lined,algoruled]{algorithm2e}
\usepackage{color}
\usepackage[title]{appendix}
\usepackage{hyperref}

\newcommand{\bfd}{\mathbf{d}}

\newcommand{\bfr}{\mathbf{r}}
\newcommand{\bfp}{\mathbf{p}}
\newcommand{\bfq}{\mathbf{q}}

\newcommand{\Kmax}{K_{\mathrm{max}}}
\newcommand{\bfgz}{\mathbf{g}_{\mathrm{z}}}

\newcommand{\bfge}{\mathbf{g}_{\mathrm{exact}}}
\newcommand{\bfgo}{\mathbf{g}_{\mathrm{obs}}}
\newcommand{\bfdo}{\bfgo}

\newcommand{\Wd}{W_{\bfd}}

\newcommand{\Rn}{\mathcal{R}^{N}}

\newcommand{\argmin}[1]{\textnormal{arg} \min_{#1}}

\begin{document}

\begin{frontmatter}

\title{IGUG: A MATLAB package for $3$D inversion of gravity data using graph theory}

\author[1]{ Saeed Vatankhah \corref{cor1}}
\ead{svatan@ut.ac.ir}
\author[2]{Vahid Ebrahimzadeh Ardestani}
\ead{ebrahimz@ut.ac.ir}
\author[3]{Susan Soodmand Niri}
\ead{susan.soodmand@ut.ac.ir}
\author[4]{Rosemary Anne Renaut}
\ead{renaut@asu.edu}
\author[5]{Hojjat Kabirzadeh}
\ead{hkabirza@ucalgary.ca}

\cortext[cor1]{Corresponding author}
\address[1]{Institute of Geophysics, University of Tehran, Tehran, Iran}
\address[2]{Institute of Geophysics, University of Tehran, Tehran, Iran}
\address[3]{Institute of Geophysics, University of Tehran, Tehran, Iran}
\address[4]{School of Mathematical and Statistical Science, Arizona State University, Tempe, AZ, USA}
\address[5]{Department of Geomatics Engineering, University of Calgary, AB, Canada}

\begin{abstract}
We present an open source MATLAB package, {\texttt{IGUG}}, for $3$D inversion of gravity data. The algorithm implemented in this package is based on methodology that was introduced by \cite{Bijani:15}. A homogeneous subsurface body is modeled by an ensemble of simple point masses. The model parameters are the Cartesian coordinates of the point masses and their total mass. The set of point masses is associated to the vertices of a weighted full graph in which the weights are computed by the Euclidean pairwise distances separating vertices.  Kruskal's algorithm  is used to solve the minimum spanning tree (MST) problem for the graph, yielding the reconstruction of the skeleton of the body described by the model parameters. The algorithm is stabilized using an equidistance function that restricts the spatial distribution of point masses and  favors a homogeneous distribution for  the subsurface structure.  The  non-linear global objective function for the model parameters  comprises the data misfit term and the stabilization equidistance function. A regularization parameter $\lambda$  is introduced to balance the two terms of the objective function, and reasonable physically-relevant bound constraints are imposed on the model parameters. A genetic algorithm  is used to minimize the bound constrained objective function for a fixed  $\lambda$, subject to the bound constraints.  A new diagnostic approach is presented for determining a suitable choice for  $\lambda$, requiring a limited number of solutions for a small set of $\lambda$.  This contrasts the use of the L-curve which was suggested for estimating the $\lambda$ in \cite{Bijani:15}.  Simulations for synthetic examples  demonstrate  the efficiency and effectiveness of the implementation of the algorithm. It is verified that the constraints on the model parameters are not restrictive, even  with less realistic bounds acceptable approximations of the body are still obtained. Included in the package is the script {\texttt{GMD.m}} which is used for generating synthetic data and for putting measurement data in the format required for the inversion implemented within \texttt{IGUG.m}. The script \texttt{Diagnostic$\_$Results.m} is included within \texttt{IGUG.m} for analyzing and visualizing the results, but can also be used as a standalone script given import of prior results.  The software can be used to verify the simulations and the analysis of real data that is presented here, The real data set uses  gravity data from the Mobrun ore body, north east of Noranda, Quebec, Canada. \end{abstract}

\end{frontmatter}
\newpage
\section{Introduction}\label{intro}
The inversion of gravity data is an efficient methodology for estimating an approximate model   of a subsurface body. Acquired gravity data on, or near, the surface are used in an automatic algorithm to estimate the defining model parameters, such as the density contrast and geometry of the subsurface body.   Using well-defined prior  information   in the inversion algorithm, an acceptable reconstruction for the subsurface is desired. Inversion methodologies include both linear and non-linear approaches, dependent on how the model is formulated.  A standard linear inversion assumes that the subsurface under the survey area is discretized as a large number of prisms of known and fixed geometry.   Then, the unknown density contrasts of each prism are estimated and   displayed to illustrate the complete geometry and density of the subsurface sources \citep{LaKu:83,LiOl:98,PoZh:99,BoCh:01,VRA:17}.   This methodology provides sufficiently useful estimates of the subsurface for high confidence mineral exploration studies.   
On the other hand, non-linear gravity inversion is usually used to find  interfaces.
For example, in hydrocarbon exploration it is important to accurately identify the depth to the basement.
Then, the geometry of the sedimentary basin is replaced with a series of juxtaposed prisms, of fixed width and known density contrast, but with unknown thickness. The shape of the sedimentry basin is obtained by estimating the thickness of each prism   \citep{Bott:60,Bl:95,ChSu:07}.  Aside from these two standard approaches, other specialized techniques have been designed to handle particular situations.  For  example, \cite{Bijani:15} developed a graph theory approach for the  $3$D inversion of gravity data in which the subsurface body is modeled as an ensemble of simple point masses. The model parameters are the Cartesian coordinates and total mass of the point masses, and the algorithm yields the reconstruction of  the skeleton of the body with the obtained coordinates and total mass. Here, as described in the following sections, we present a MATLAB package to implement gravity inversion  based on  some extensions of the graph theory approach of \cite{Bijani:15}. 

It is well-known using the theory of Green's equivalent layer, that the solution of the gravity inverse problem is non-unique \citep{LaNa:12}. Moreover, the gravity data measurements are always contaminated by noise due to both instrumental errors and modeling simplifications. Thus, in obtaining a  geologically plausible solution given the measured data, prior information has to be incorporated into the solution process. A stabilizing regularization term is  imposed to assure that the solution is not overly contaminated by noise in the data, and biases the search space for the model parameters to a space  defined by the interpreter. For example, as used in linear inversion, $L_{0}$ and $L_{1}$ norm stabilizers lead to the reconstruction of sparse solutions \citep{LaKu:83,PoZh:99,BoCh:01,VAR:15,VRA:17}, a depth weighting function reduces the impact of  the natural decay of the sensitivity matrix with depth \citep{,LiOl:98}, and imposed $L_2$ norm stabilization with a derivative operator provides  smooth solutions \citep{,LiOl:98}. Non-linear inversions have been stabilized by constraining the  density variation with depth \citep{ChSu:07} and applying a total variation regularization \citep{Martins:11}.  In the graph theory approach    of \cite{Bijani:15}   the equidistance function stabilization was introduced.  The set of point masses are associated to the vertices of a weighted full graph  in which the weights between pairwise vertices are computed from the Euclidean distances between the vertex pairs.    Kruskal's algorithm is used to solve the minimum spanning tree (MST) problem for the graph, and the equidistance function is computed using the MST. This function restricts the spatial distribution of the point masses and thus provides a solution that prefers a homogeneous spatial distribution.  Consequently, a skeleton of the body is reconstructed.     We note that it is also possible to include prior information on the model parameters so that physically realistic bound constraints, determined by knowledge of the local geology, are imposed. 

General gravity inversion incorporating stabilization requires the minimization of an objective function comprising the data misfit term and the stabilizing  function with balancing provided by a regularization parameter, $\lambda$.  Deterministic  algorithms for the optimization, such as Levenberg-Marquardt or Gauss-Newton,  require the use of derivative information of the objective function, and find  the minimum of the non-linear objective function. They will not, however, necessarily distinguish between global and local minima, \citep{ZePo:93}. Convergence to a local minimum is likely and is particularly dependent on the initial model.  As an alternative, optimization based on a controlled random search can be used \citep{Montana:94}. Algorithms in this class, such as simulated annealing and natural genetic selection, simulate naturally-occurring phenomena and do not require any derivative information for the objective function. Here, we chose to use the genetic algorithm (GA) which employs a  random search algorithm based on the mechanisms of natural selection and natural genetics. 

\textit{Overview of main scientific contributions.} \cite{Bijani:15} introduced the use of graph theory for the three dimensional inversion of gravity data.  Our approach implements and extends the algorithm. 
(i)  Weighting of the data misfit term is introduced using  knowledge of the noise in the measured data. (ii) An effective technique for  determining $\lambda$ based on a linear regression (data fitting)  analysis of the convergence curves for the equidistance stabilizing function with a statistical measurement of the reliability of the data fitting is presented.   (iii) The inversion algorithm  is available as open source MATLAB code  and provides multiple options for picking the parameters of the GA. (iv) An accompanying script for  generating a synthetic model is provided. This work, therefore, realizes the original proposal of \cite{Bijani:15} as a tool for the general inversion of three dimensional gravity data. 
The algorithm is open source and available at  \url{https://math.la.asu.edu/~rosie/research/gravity.html}, along with a full description of the algorithm implementation and example simulations. 

The paper is organized as follows. In Section~\ref{forwardmodel} we present the forward model for  the gravity data, leading immediately to the inversion formulation to be solved using the GA, as described in Section~\ref{inversion}.  The specific GA is presented algorithmically in Algorithm~\ref{algorithm1} and necessary components of the graph theory are also provided. Section~\ref{code} describes how  the presented Matlab software can be used to both  generate data and perform the inversion. The use of the software is illustrated in Section~\ref{synthetic},  with a discussion of regression analysis to find $\lambda$ in Section~\ref{diagnostics}. Finally, in Section~\ref{real}, results are presented for the application of the method  on gravity data from the Mobrun ore body, north east of Noranda, Quebec, Canada.

\section{Inversion methodology}
In this section we briefly review the gravity inversion based on graph theory. For more details the readers should refer to \cite{Bijani:15}. 

\subsection{The forward model}\label{forwardmodel}
\begin{figure}[H]
\centering
\subfloat{\includegraphics[width=0.6\textwidth]{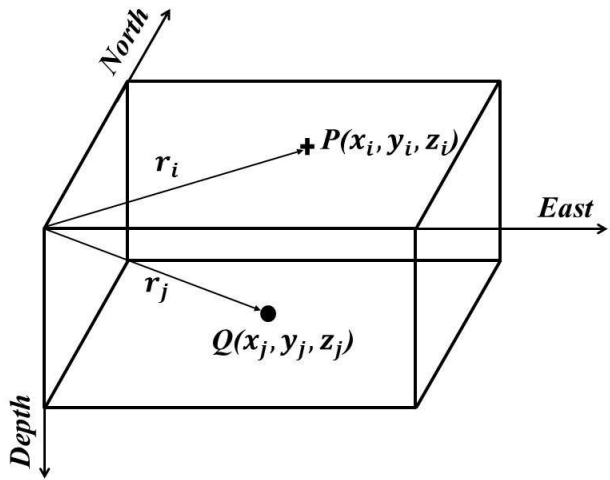}}
\caption{A single point mass located in the subsurface at point $Q$ which has coordinates $(x_j,y_j,z_j)$. Point $P$ is the gravity station located at the surface with coordinates $(x_i,y_i,z_i)$. The vectors $\bfr_i$ and $\bfr_j$ are the position vectors of $P$ and $Q$, relative to the origin, respectively.}\label{fig1}
\end{figure}

Suppose a point mass  in the subsurface is located at point $Q$   and has coordinates $(x_j,y_j,z_j)$, Fig.~\ref{fig1}. The resulting vertical component of the gravity field at point $P$ on the surface with coordinates $(x_i,y_i,z_i)$  is  given by, \citep{Bl:95},
\begin{align}\label{forward}
g_z(\bfr_i,\bfr_j) = -\gamma  \frac{m_j(z_i-z_j)}{\| \bfr_i-\bfr_j\|_2^3}.
\end{align}
Here $\gamma$ is the universal gravity constant, $m_j$ is the value of the mass assigned to point $Q$,   vectors $\bfr_i$ and $\bfr_j$ are respectively the position vectors of $P$ and $Q$ relative to the origin, and  $\|.\|_2$ indicates the Euclidean norm of a vector. The total vertical gravity component at point $P$ due to  $M$ point masses in the subsurface obtained by superposition over all point masses is given by
\begin{align}\label{superposition}
(\bfgz)_i = \sum_{j=1}^{M} g_z(\bfr_i,\bfr_j).
\end{align}   
Here $(\bfgz)_i$ denotes the $i^{\mathrm{th}}$ component of the vector $\bfgz  \in \Rn$ which comprises the responses at all stations $i=1:N$ on the surface, and describes the forward gravity model. Inversion of the model requires the estimation of the point masses and their positions given the measurements of the gravity anomaly at the $N$ gravity stations. The estimated set of  point masses indicates a skeleton of the geometry and  provides the total mass of the causative  subsurface source relative to the background mass of the surrounding area,  \citep{Bijani:15}.  
\subsection{The inverse model}\label{inversion}
Suppose that the observed gravity data for a homogeneous source are given by the components of the vector  $\bfgo \in \Rn$ and that the point masses, randomly spread throughout the domain,  have the same mass, $m_j=m_p$ for all $j$. Then the total mass is assumed to be $m_t=Mm_p$. Suppose  that the Cartesian coordinates of the sources are assigned to vector $\bfp \in \mathcal{R}^{3M}$ ordered as
\begin{align}\label{pvector}
\bfp = (x_1,y_1,z_1,\cdots,x_M,y_M,z_M)^T, 
\end{align}
and that the resulting vector of model parameters of dimension $3M+1$ is given by 
\begin{align}\label{qvector}
\bfq = (m_t,\bfp^T)^T.
\end{align}
It is desired to find vector $\bfq$ which generates forward vector $\bfgz(\bfq)$ that predicts the  observed gravity vector $\bfgo$ at  the given noise level.  The data fitting constraint is imposed using the data-misfit term 
\begin{align}\label{datafit}
\Phi(\bfq) = \| \Wd(\bfgo-\bfgz(\bfq)) \|_2^2, 
\end{align}
for diagonal data weighting matrix $\Wd$, with entries $(\Wd)_{ii}=\sigma_i^{-1}$ where $\sigma^2_i$  is the assumed variance of the error in the $i^{\mathrm{th}}$ measurement  $(\bfgo)_i$. Equivalently it is assumed that the  noise in the data is Gaussian and uncorrelated, and $\Wd$ is the inverse square root of the diagonal covariance matrix for the noise.

The non-uniqueness of the gravity inversion problem, and the associated sensitivity of the solution to noise in the data, requires that the set of potential solutions $\bfq$ that minimize $\Phi(\bfq)$  is reduced by the introduction of a stabilization term in the minimization. \cite{Bijani:15} introduced the use of concepts from graph theory for stabilizing the solution of \eqref{datafit}.  First suppose that the point masses are considered as vertices of a full\footnote{For a full graph all vertices are connected.} graph with the edges between the vertices connecting all the point masses. For a weighted full graph each edge between vertices $i$ and $j$ is assigned a weight $d_{ij}$. In this case, $d_{ij}$ is the Euclidean distance  between point masses $i$ and $j$. Thus closer points have a smaller weight.  Imposing $M$ point masses, the minimum spanning tree (MST) problem finds the graph that connects all point masses while minimizing the total distance in the graph, namely it forms the least distance spanning  tree (LDST) for the graph. The minimum number of edges of the LDST for $M$ point masses is $M-1$.   Kruskal's algorithm, \citep{Krus:56},  is a greedy algorithm for finding the subset of edges that form the LDST. We use   $\mathbf{d}^{MST}(\bfp) \in \mathcal{R}^{M-1}$ to denote the vector  containing the  lengths of all  edges  of the LDST, and  $\overline{\mathbf{d}}^{MST}(\bfp)$ to be the mean of the distances of the MST.  Then, as a  further stabilization of the search space, \cite{Bijani:15} constrained the MST to have edges of equal length yielding the stabilizing equidistance function
\begin{align}\label{equidis}
\Theta(\bfp) =\sum_{i=1}^{M-1}[\mathbf{d}_{i}^{MST}(\bfp)-\overline{\mathbf{d}}^{MST}(\bfp)]^2,
\end{align} 
where $\mathbf{d}_{i}^{MST}(\bfp)$ contains the lengths $d_{ij}$ for $j\ne i$.
Here   $\Theta(\bfp)$   effectively minimizes  the variance in the edge lengths against their average and thus biases the solution toward a homogeneous $3$D spatial distribution of point sources in the subsurface. Consequently, the inversion algorithm is able to reconstruct the skeleton of the subsurface body. 

Given the data misfit function $\Phi$ and the stabilization term $\Theta$,   a balancing parameter, or regularization parameter, $\lambda$, is introduced. This trades off the relative importance of the data misfit and stabilization terms in the objective function
\begin{align}\label{globalobjective}
\Gamma(\bfq) =\Phi(\bfq)+\lambda \Theta(\bfp).
\end{align} 
An algorithm is  required to obtain $\bfq_{\mathrm{opt}}$ that  minimizes $\Gamma$ for a fixed $\lambda$. Further, an approach is required that efficiently selects a $\lambda$ which generates an acceptable solution given the measured data.  

\cite{Bijani:15} suggested using a GA for the minimization of $\Gamma(\bfq)$,  e.g. \citep{GoHo:88,Montana:94}. The method starts from an initial random population, consisting of a number of individuals $\bfq$, and iteratively improves the estimated solution. Constraints on the model parameters (Cartesian coordinates and total mass) are  used in all stages of the GA, allowing the inclusion of prior information on the model. At each iteration  each individual of the  population is given a fitness (i.e., a value of the objective function $\Gamma(\bfq)$). The fittest individuals are selected for reproduction in order to produce offspring that augment the set of individuals used for the next generation at step $k+1$.     The individuals with highest fitness are paired and reproduced through a crossover operation, giving birth to an offspring population. A small percentage of this new population is arbitrarily mutated, dependent on a given mutation rate, so different areas of the search space can be explored.  This assists with avoiding local minima in the optimization process.  The new population is also evaluated, allowing only the fittest individuals to survive, and the process is repeated. The GA  terminates when either the solution satisfies the noise level, 
\begin{equation}\label{chitest}\Phi( \bfq )= \| \Wd(\bfgo-\bfgz(\bfq)) \|_2^2 \leq N+\sqrt{2N},\end{equation} or a certain number of generations, $\Kmax$, is reached. The best individual of all generations is selected as the final estimate,  $\bfq_{\mathrm{opt}}$. The inversion  methodology for a fixed $\lambda$ is summarized in Algorithm~\ref{algorithm1}.

\begin{algorithm}
\caption{\texttt{IGUG:} Minimization of $\Gamma(\bfq)$ for gravity inversion using a genetic algorithm, given measured data $\bfgo$ and estimated noise distribution on the data via $\Wd$.}\label{algorithm1}
\begin{algorithmic}[1]
\REQUIRE Genetic algorithm parameters as detailed in Table~\ref{Tab:OptimParam}. 
\FOR{$\ell=1$ \textbf{to} $\mathrm{noq}$}  
\STATE { Generate  random population $\bfq^{(\ell)}$. Impose coordinate and mass constraints:  $x_{{\mathrm{min}}} \le x_j\le x_{{\mathrm{max}}}$, $y_{{\mathrm{min}}} \le y_j\le y_{{\mathrm{max}}}$, $z_{{\mathrm{min}}} \le z_j\le z_{{\mathrm{max}}}$, $m_{t_{\mathrm{min}}}\le m_t\le m_{t_{\mathrm{max}}}$.} \ENDFOR
\STATE $k=0$. $\Phi(\bfq_{\mathrm{opt}}) = 10^6$.
\WHILE{($k<\Kmax$) \textbf{\&}  ($\Phi(\bfq_{\mathrm{opt}})>N+\sqrt{2N}$)}
\FOR{$\ell=1$ \textbf{to}  $\mathrm{noq}$}  \STATE {Generate a full graph for $\bfq^{(\ell)}$. Use Kruskal's algorithm to find the  least distance spanning tree for $\bfq^{(\ell)}$. Calculate $\mathbf{d}_{i}^{MST}(\bfq^{(\ell)})$ and $\overline{\mathbf{d}}^{MST}(\bfq^{(\ell)})$.
Compute $ \Gamma(\bfq^{(\ell)}) =\Phi(\bfq^{(\ell)})+\lambda \Theta(\bfp^{\ell})$}.
 \ENDFOR 
\STATE $\bfq_{\mathrm{opt}}=\argmin{\ell}\Gamma(\bfq^{(l)})$.
\STATE Use GA to generate  new population via genetic selection, mutation and crossover. Impose constraints at all stages.
\ENDWHILE
\ENSURE $\bfq_{\mathrm{opt}}$ and  iteration count $k$.
\end{algorithmic}
\end{algorithm}

\section{Software Package}\label{code}
The software consists of three main scripts. 
 \begin{description}
\item[\href{http://math.la.asu.edu/~rosie/research/gravity/htmlfiles/GMD.html}{\texttt{GMD.m}}]  is used to generates a synthetic model and its gravity data subject to a user-specified survey area and subsurface geometry.  It can also be used to create the appropriate real data set for inversion, using the measured data, noise distribution and survey area. 
\item[\href{http://math.la.asu.edu/~rosie/research/gravity/htmlfiles/IGUG.html}{\texttt{IGUG.m}}]  loads the data file produced for either synthetic or real data and performs the inversion to find $\bfq_{\mathrm{opt}}$. It can be run for a single $\lambda$, or a range of values for $\lambda$.
\item[\href{http://math.la.asu.edu/~rosie/research/gravity/htmlfiles/Diagnostic_Results.html}{\texttt{Diagnostic$\_$Results.m}}] is used to analyze the results and provides an approach for determining $\lambda$. It is included at the end of \texttt{IGUG.m} and is also a standalone script for analyzing output from \texttt{IGUG.m}. 
\end{description}
Extensive discussion on each script is available in the \href{http://math.la.asu.edu/~rosie/research/gravity/IGUGDescription.pdf}{documentation}, including specifics on the input and output parameters.  This information also discusses the directory structure and provides examples of the usage of the \href{http://math.la.asu.edu/~rosie/research/gravity.html}{package}. We review the important components of these main scripts below.

\subsection{{\texttt{GMD.m}}}\label{gmd}
\href{http://math.la.asu.edu/~rosie/research/gravity/htmlfiles/GMD.html}{\texttt{GMD.m}} is an easy to use MATLAB code for producing the vertical component of the gravity field, the data vector $\bfgo$,  for a user defined synthetic model at a specified noise level. The model is generated   using an ensemble of one or more prisms. For example, a vertical dike may need just one prism, but a more complex geometry is represented by a set of prisms. The parameters of the simulation, including the survey volume, subsurface geometry, noise variance for $\Wd$ and all parameters required for the inversion  are saved for import to the inversion module \texttt{IGUG.m}. \texttt{GMD.m} can be edited by the user for more general usage when generating synthetic data sets, and in particular to modify the model for the noise. 

{\texttt{GMD.m} is also used to read a real data file with the measured data set that includes the data vector $\bfgo$, an estimate for $\Wd$ and the coordinates for the locations of the stations. In this case the user is asked to provide the additional parameters that are required for the inversion, including the  survey volume and the parameters required for the inversion, but does not assume any knowledge of the subsurface geometry. 

For both synthetic and real data sets {\texttt{GMD.m} provides a plot of the survey volume and the gravity anomaly, and in the case of synthetic data the subsurface geometry is inset within the survey volume. This allows the user to check that the information has been correctly provided. The  outline for {\texttt{GMD.m} when used for synthetic data sets is provided in Algorithm~\ref{algorithm2}.  A simple modification is coded for the case with real data. 

\begin{algorithm}
\caption{\texttt{GMD:} Generating a synthetic model: In all cases default values may be chosen.}\label{algorithm2}
\begin{algorithmic}[1]
\REQUIRE Initial exact gravity data is empty. $\bfge =\lbrack  \rbrack$
\STATE \textit{Generate the survey domain:} Provide coordinates of the origin,  extension of the volume in East, North and depth dimensions.
\STATE \textit{Data for generating the anomaly:} Give distances between stations in East and North directions and number of prisms $\mathrm{noc}$ used for the subsurface structure. Pick a noise level index: $j$.
\FOR{$k=1$ {\bf to} $\mathrm{noc}$}
\STATE {\textit{Define the substructure} Give the three dimensional coordinates and the density of prism $k$. }
\STATE{\textit{Generate the gravity anomaly for prism $k$:}  $\bfge^{(k)}$ }
\STATE{\textit{Accumulate exact gravity:}  $\bfge= \bfge+\bfge^{(k)}$} 
\ENDFOR
\STATE \textit{{Generate noisy gravity anomaly and provide noise distribution:}}   $\bfgo$ and $\Wd$.  
\STATE \textit{{Check data input:}} Plot true and noisy data and the subsurface geometry.
\ENSURE Save parameters $\bfgo$,  $\Wd$, discretization choices, and survey area descriptions to \texttt{DataNj.mat}.
\end{algorithmic}
\end{algorithm}

\subsection{\texttt{IGUG.m}}\label{igug}
\href{http://math.la.asu.edu/~rosie/research/gravity/htmlfiles/IGUG.html}{\texttt{IGUG.m}} implements the inversion methodology based on Algorithm~\ref{algorithm1}. It requires a synthetic data set such as produced using \texttt{GMD.m} or can be used for real data with the same format, potentially also generated using \texttt{GMD.m} as noted in Section~\ref{gmd}. Parameters for the GA must also be given, as indicated in Table~\ref{Tab:OptimParam}. The constraint conditions on  the horizontal coordinates  can be  defined by analyzing the amplitude of the observed data. The constraints for the total mass and the depth coordinates can be determined from prior information. Our experience indicates that it is not necessary to determine  tight constraints. Thus, when no prior information is available wide constraints still provide acceptable results.  It is possible to use all  parameters of the GA set to default values, but the user is interrogated as to whether values should be altered.  

\subsection{\texttt{Diagnostic$\_$Results.m}}\label{sec:DR}
\href{http://math.la.asu.edu/~rosie/research/gravity/htmlfiles/Diagnostic_Results.html}{\texttt{Diagnostic$\_$Results.m}} can be used to assist in interpretation of the results of the genetic algorithm and to select the parameter $\lambda$. The user has the option to  plot obtained results for visual inspection without any further analysis, if all dialogue boxes are answered with ``No".  In this case plots are given of $(k,\Gamma(k))$,  $(k,\Phi(k))$ and $(k, \log(\Theta(k))$ for each choice of $\lambda$ and  the resulting point mass distribution will be provided within the survey volume. A table of results that summarizes the final values of $k$, $\Gamma$, $\Phi/(N+\sqrt{2N})$ and $\Theta$ for the given $\lambda$ is displayed in the command window. 

Selecting ``Yes"  for linear regression analysis introduces a quantitative diagnostic for the analysis of the results based on fitting   the convergence curves to a straight line. Briefly, given $n$ data points $(x_i,y_i)$ we seek the linear approximation $y(x)=ax+b$ by minimizing 
\begin{equation}\label{eq.ls}F(a,b)=\sum_{i=1}^n \left(y_i - (a  x_i +b)\right)^2\end{equation}  for which the solution is immediately available  in terms of the mean values of $x$, $y$, $xy$, and $x^2$
\begin{align*}
a&= \frac{\overline{xy} -(\overline{x})(\overline{y})}{\overline{x^2}-\overline{x}^2}, \quad  b=\overline{y}-a\overline{x},
 \end{align*}
 where   $\overline{ \cdot}$ denotes the mean value. 
 Moreover, denoting the predicted values given by $\hat{y}_i=ax_i+b$, the $R^2$ statistic, or coefficient of determination,  which is a measure of how well the linear model predicts the data, is available as 
 $$R^2 =  \left(\frac{\sum_{i=1}^n ( \hat{y}_i  - \overline{y})}{\sum_{i=1}^n (y_i  - \overline{y})}\right)^2. $$
When $R^2$ is close to $1$ we deduce a good prediction is achieved, but close to $0$ we deduce that the line is not a good predictor of the data.  As we will see from the data, when $\lambda$ is too small, instability in the convergence of $\Theta$ with increasing $k$ is indicative of  a solution that is under-regularized, or that the solution is not progressing and $\Theta$ is at the noise level for the computation. This can be assessed applying the regression analysis. Thus, for the diagnostics we present the option for regression
analysis (data fitting) for $\Gamma$, $\Phi$ and $\log(\Theta)$ as  function of $k$.  The linear regression results are then also illustrated in the plots and given in the table of results. We will show how these results can be used to efficiently estimate an appropriate choice for $\lambda$ at limited cost. Finally there is the option to save all figures   in \textrm{.jpg} format,  and to export the table of results to a spread sheet.  

\section{Results}
We present results using the software package for the inversion of both simulated and real data sets, Sections~\ref{synthetic}-\ref{diagnostics} and \ref{real}, respectively. All reports on timing are presented for an implementation using MATLAB Version 9.4.0.813654 (R2018a) running under the Mac OS X  Version: 10.13.6 Operating System. These results can be replicated using the simulated and real data sets  \texttt{DataN4.mat} and \texttt{AllRealData.mat} that are provided with  \href{http://math.la.asu.edu/~rosie/research/gravity/Codes.zip}{the codes,}
 but it should be noted that all results depend on randomization in the GA and thus obtained results will be equivalent but not exact replications. 
 \subsection{Synthetic example}\label{synthetic}
We consider the  example of a dipping dike model, Fig.~\ref{fig2}. \texttt{GMD.m} was used to generate the model for a dike with  three prisms, $\mathrm{noc}=3$. The dimensions of the prisms are given in Table~\ref{tab1}. The density contrast of the dike is $1$~$g/cm^3$ and its total mass is $108 \times 10^9$ $kg$. Gravity data of the model, $\bfge$, were generated on the surface for a grid of $41 \times 31 =1271$ points with grid spacing $50$~m. Gaussian noise with standard deviation $( 0.02 (\bfge)_i + 0.001 \| \bfge \|_2)$ is added to each datum yielding the noisy data set, $\bfdo$, illustrated in Fig.~\ref{fig3}. The selected parameters for performing the inversion are given in Table~\ref{tab2} and a summary of the results is provided in Table~\ref{tab3}.  
\begin{figure}[H]
\subfloat{\label{fig2a}\includegraphics[width=0.5\textwidth]{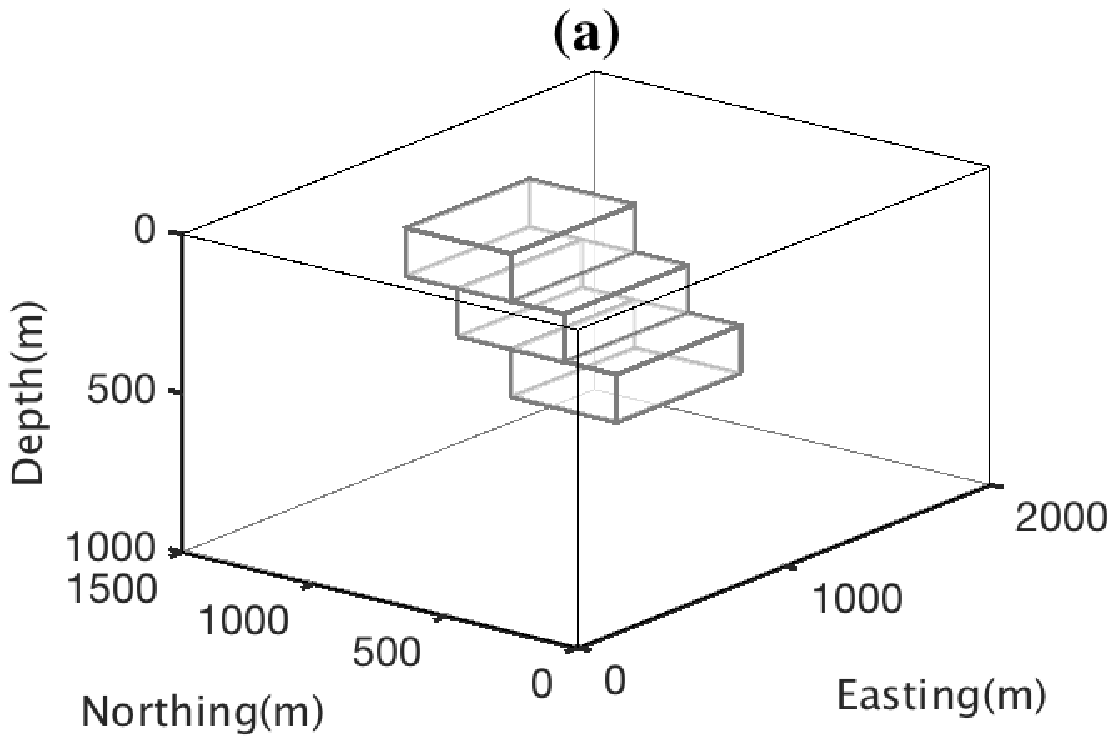}}
\subfloat{\label{fig2b}\includegraphics[width=0.5\textwidth]{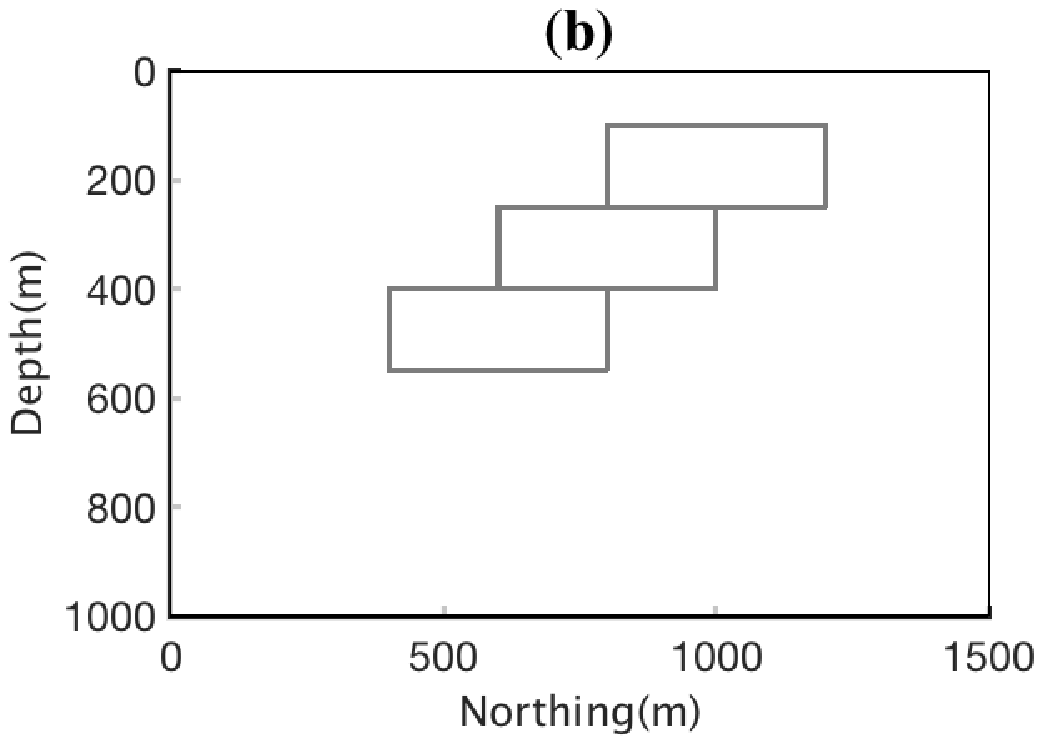}}
\caption{Model of a dipping dike with density contrast of $1$~$g/cm^3$. (a) A perspective view of the model; (b) The cross-sectional view of the model.}\label{fig2}
\end{figure}

\begin{figure}[H]
\centering
\subfloat{\label{fig3a}\includegraphics[width=0.47\textwidth]{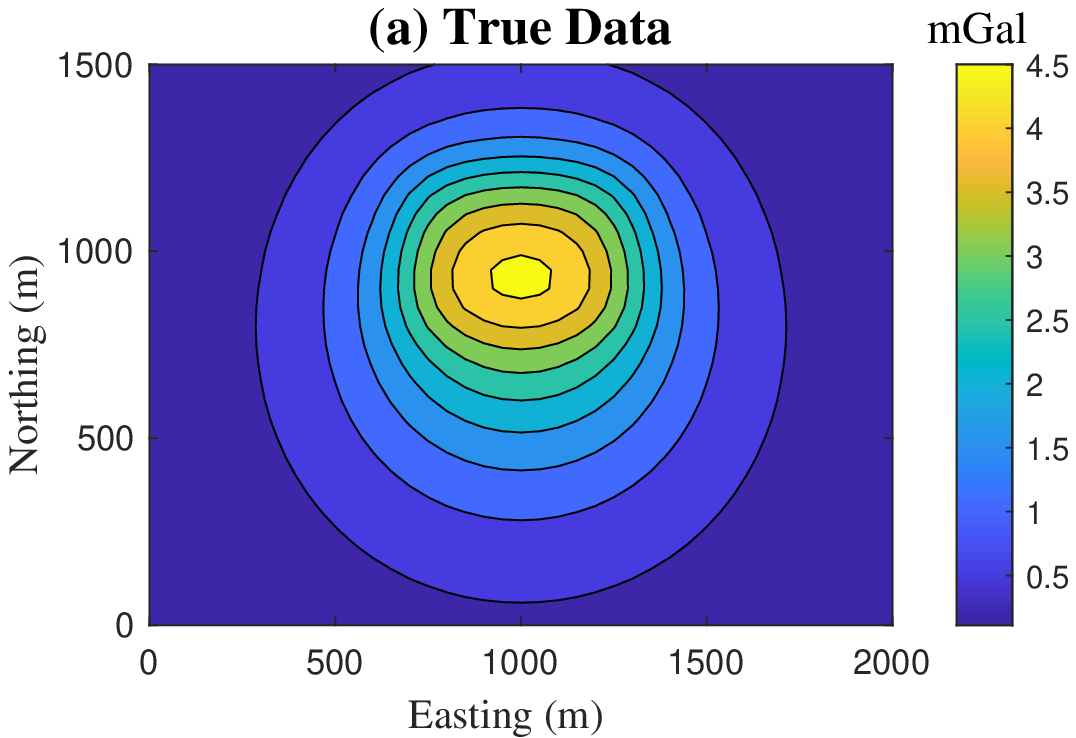}} 
\subfloat{\label{fig3b}\includegraphics[width=0.47\textwidth]{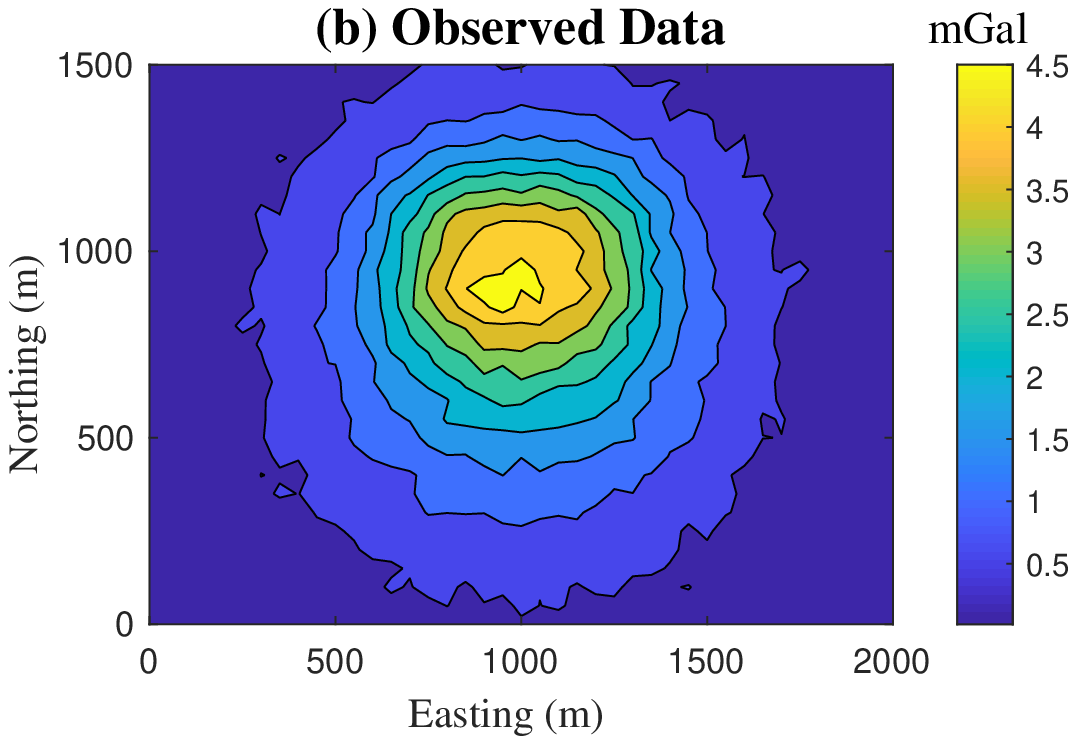}}

\caption{The gravity anomaly produced by the  model shown in Fig.~\ref{fig2}, without noise in Fig.~\ref{fig3a} and contaminated by  noise  with   $\sigma_i=( 0.02 (\bfge)_i + 0.001 \| \bfge \|_2)$ in Fig.~\ref{fig3b}.}\label{fig3}
\end{figure}

\begin{table} [H]
\scriptsize
\caption{The dimensions of the prisms used to form the model in Fig.~\ref{fig2}.}
\centering
\begin{tabular}{|c|c|c|c|}
\hline
Prism  & East (m) & North (m)& Depth \\
\hline
Upper  & $700$ to $1300$   & $800$ to $1200$  & $100$ to $250$ \\
Middle & $700$ to $1300$   & $600$ to $1000$  & $250$ to $400$ \\
Lower  & $700$ to $1300$   & $400$ to $800$   & $400$ to $550$\\
\hline 
\end{tabular}
\label{tab1}
\end{table}

\begin{table} [H]
\scriptsize
\caption{Parameters used as input of Algorithm~\ref{algorithm1} to perform inversion for data of Fig.~\ref{fig3}. Coordinates are given in meters, $m$, and mass in kilograms, $kg$.}
\centering
\begin{tabular}{|c|c|c|c|c|c|c|c|c|c|c|}
\hline
  $\mathrm{noq}$ & $x_{{\mathrm{min}}}$ & $x_{{\mathrm{max}}} $& $y_{{\mathrm{min}}} $& $y_{{\mathrm{max}}} $& $z_{{\mathrm{min}}} $& $z_{{\mathrm{max}}}$ & $m_{t_{\mathrm{min}}} $& $m_{t_{\mathrm{max}}} $ \\
\hline
  $100$& $400$& $1600$& $100$ & $1400$& $20$& $1000$ & $70e9$ & $150e9$ \\
\hline 
\end{tabular}
\label{tab2}
\end{table}

First we contrast the results for $\lambda = [100, .1, .00001]$ with fixed $\Kmax=200$ and $20$ point masses, in  Figs.~\ref{fig4}, \ref{fig5}, and \ref{fig6}, respectively. In each case   we  illustrate the reconstructed model superimposed on the dipping dike structure,  the convergence of the equidistance function $\Theta$ as a function of the iteration number $k$,   and the data predicted by the model.    Comparing the point mass distribution we see that for over regularization, large $\lambda$, we see a dispersed set of points that  does not approximate a skeleton of the original body (Figs.~\ref{fig4a} and \ref{fig4b}),  greater homogeneity is achieved for $\lambda=.1$  (Figs.~\ref{fig5a} and \ref{fig5b}), and that there are a greater number of points not connected to the body for the under regularized case with very small $\lambda=.00001$ (Figs.~\ref{fig6a} and \ref{fig6b}). The progression of $\Theta$, illustrated in Figs.~\ref{fig4c}, ~\ref{fig5c} and ~\ref{fig6c}, shows a distinct difference in the convergence behavior, showing instability as $k$ increases for $\lambda=.00001$. Contrasting the predicted anomalies, Figs.~\ref{fig4d}, ~\ref{fig5d} and ~\ref{fig6d}  with Fig.~\ref{fig3a},  it is clear that the over regularized result does not yield a good approximation. Moreover, considering the quantitative results in Table~\ref{tab3},  for over-regularization the total mass is  under estimated and the final value of   $\Phi$  is also significantly larger than the desired noise level.   From these results, we conclude that while the final value of $\Phi$ is closer to the desired estimate $ N+\sqrt{2N}\approx 1321$ for $\lambda=.00001$, the lack of stability in the estimate of $\Theta$ with $k$, as indicated by the low $R^2$ value,  is suggestive that the convergence is not stable, and that this result would be less reliable than the choice with $\lambda=.1$. It should be noted that the costs are effectively independent of $\lambda$, all timings are on the order of $120$ seconds, for the determination of the solution, with fixed $M=20$ and $\Kmax=200$.
\begin{table} [H]
\scriptsize
\caption{The results of the inversion for the given selections of $\lambda$, $M$ and $\Kmax$. In all cases $\Phi(\bfq_{\mathrm{opt}})> N+\sqrt{2N} \approx 1321$ at the final iteration.  }
\centering
\begin{tabular}{|c|c|c|c|c|c|c|c|}
\hline
Figure&$\lambda $ &$M$&$\Phi(\bfq_{\mathrm{opt}})$& $m_{t} (kg)$ &$\Kmax$ & Time (seconds)& $R^2$\\
\hline
4&$100$  &$20$& $63209$   & $89.7e9$  & $200$&$117.6$&$.9883$ \\
5&$.1$ &$20$& $2254$   & $119.7e9$ & $200$& $121.1$ &$.8942$\\
6&$.00001$  &$20$& $1483$   &  $115.6e9$ & $200$&$117.2$&$.3377 $\\  
7&$.1$ &$20$&$1609$& $115.4e9 $   & $1000 $ & $589.9$&$.5949$ \\
8&$.1$ &$40$&$3372$&$130.2e9$  & $ 200$ & $176.4$&$.8874$ \\ 
\hline 
\end{tabular}
\label{tab3}
\end{table}
In Fig.~\ref{fig7} equivalent results are illustrated using $\lambda=.1$ but by increasing $\Kmax$ to $1000$. A slightly more compact solution is obtained, $\Phi$ is closer to the desired value, and a slightly  better estimate of the mass is achieved, but the cost has increased almost linearly to $589.9$ seconds. Thus increasing $\Kmax$ may achieve an improved solution, but the noise level is still not achieved and the extra cost may not be desirable. We also note that the decrease in $\Theta$ levels out for increasing $k$. Finally, we examine the impact of increasing the number of point masses to $M=40$, as illustrated in Fig.~\ref{fig8}. The results are not noticeably improved compared to the case with $M=20$ and the time is  increased by about $55$ seconds, $33\%$ more expensive. The cost does not increase linearly with $M$ and it may be possible to obtain some improvement in  results for some runs of the GA.

From the presented results, we conclude that when (i) there is a small data misfit $\Phi$ and when (ii)  $\Theta(\bfp)$ exhibits stable convergence, the solution is neither over or under regularized, and the solution with the given $\lambda$  the reconstructed point masses provide a good  approximation of the the skeleton of homogeneous source.  Thus, in general, the optimum parameter can be estimated without running the code for a large number of  values of $\lambda$, as is required for example with the time-consuming L-curve approach suggested by \citep{Bijani:15}.

\begin{figure}[H]
\subfloat{\label{fig4a}\includegraphics[width=0.25\textwidth]{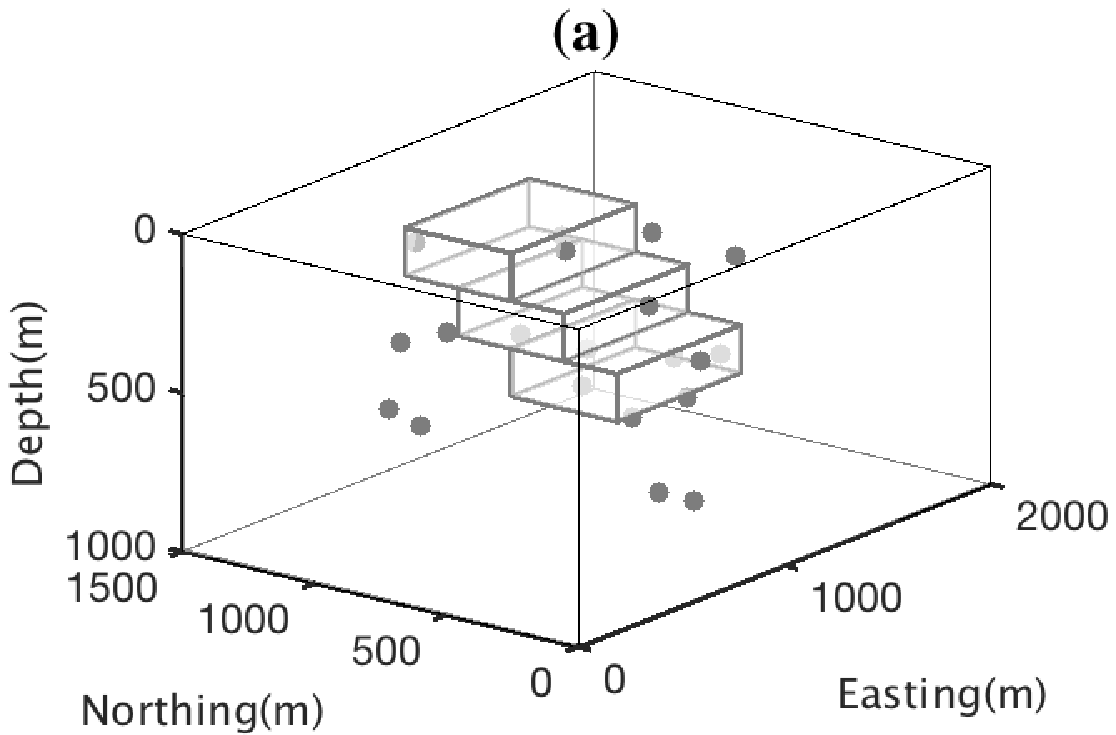}}
\subfloat{\label{fig4b}\includegraphics[width=0.25\textwidth]{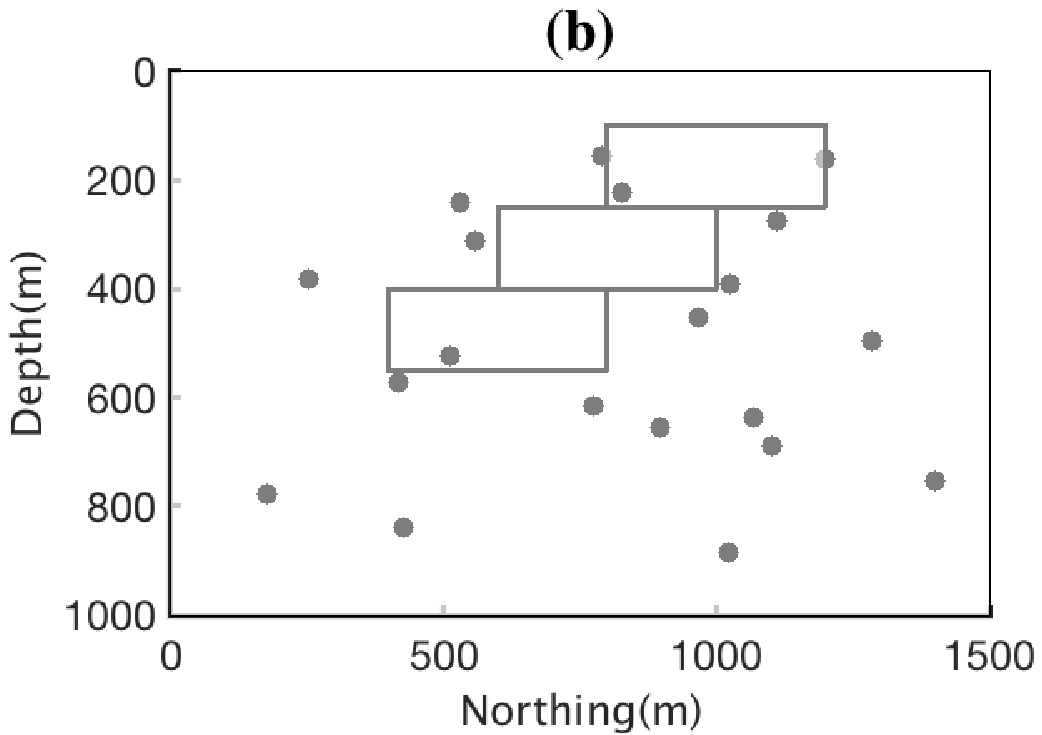}} 
\subfloat{\label{fig4c}\includegraphics[width=0.25\textwidth]{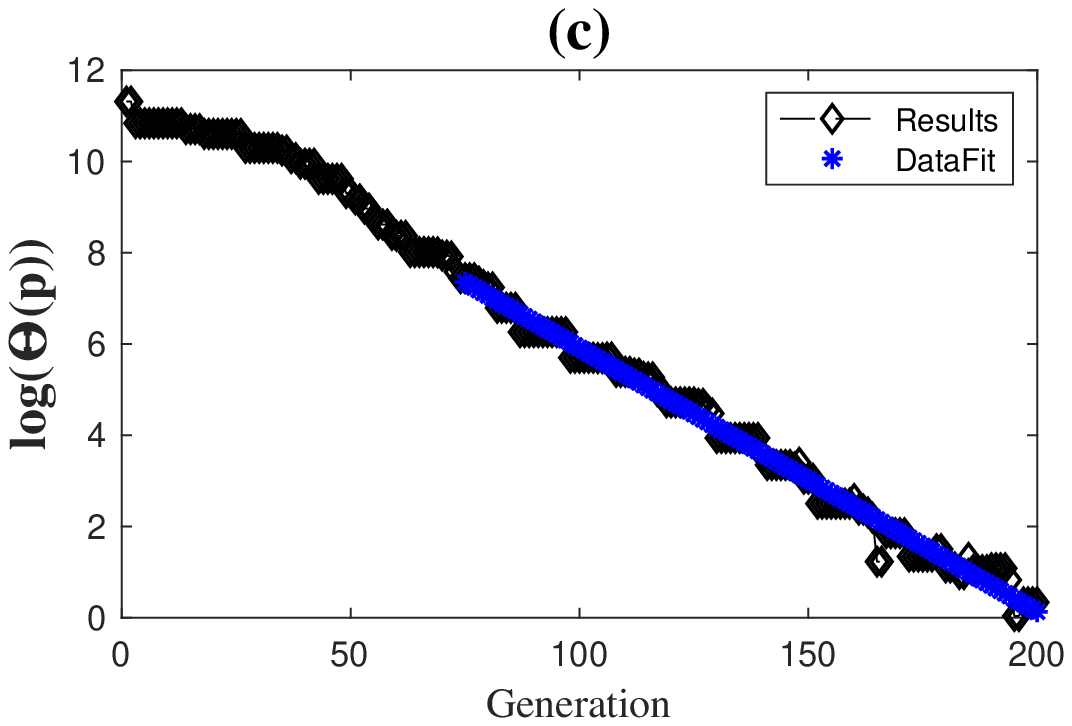}}
\subfloat{\label{fig4d}\includegraphics[width=0.25\textwidth]{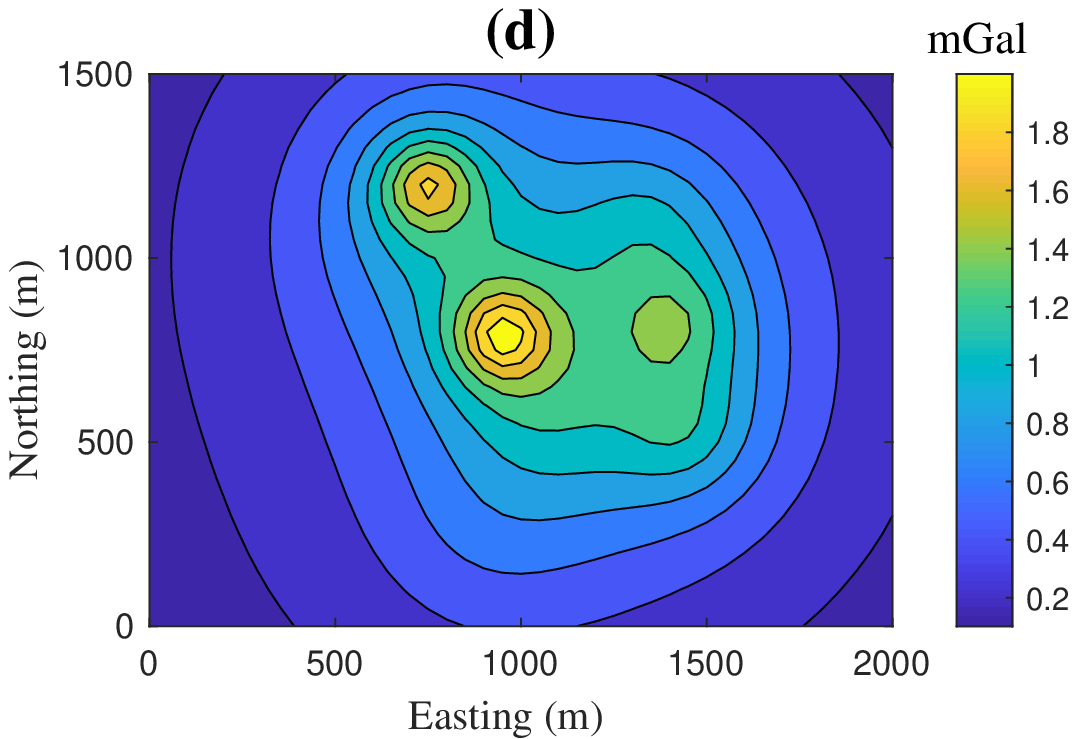}}
\caption{Inversion results using Algorithm~\ref{algorithm1} with regularization parameter $1 \times 10^{2}$, $\Kmax=200$ and $M=20$. (a) A perspective view of the point masses; (b) The cross-sectional view of the point masses; (c) The equidistance function for the best solution at each iteration; (d) The data predicted  by the reconstructed model.}\label{fig4}
\end{figure}

\begin{figure}[H]
\subfloat{\label{fig5a}\includegraphics[width=0.25\textwidth]{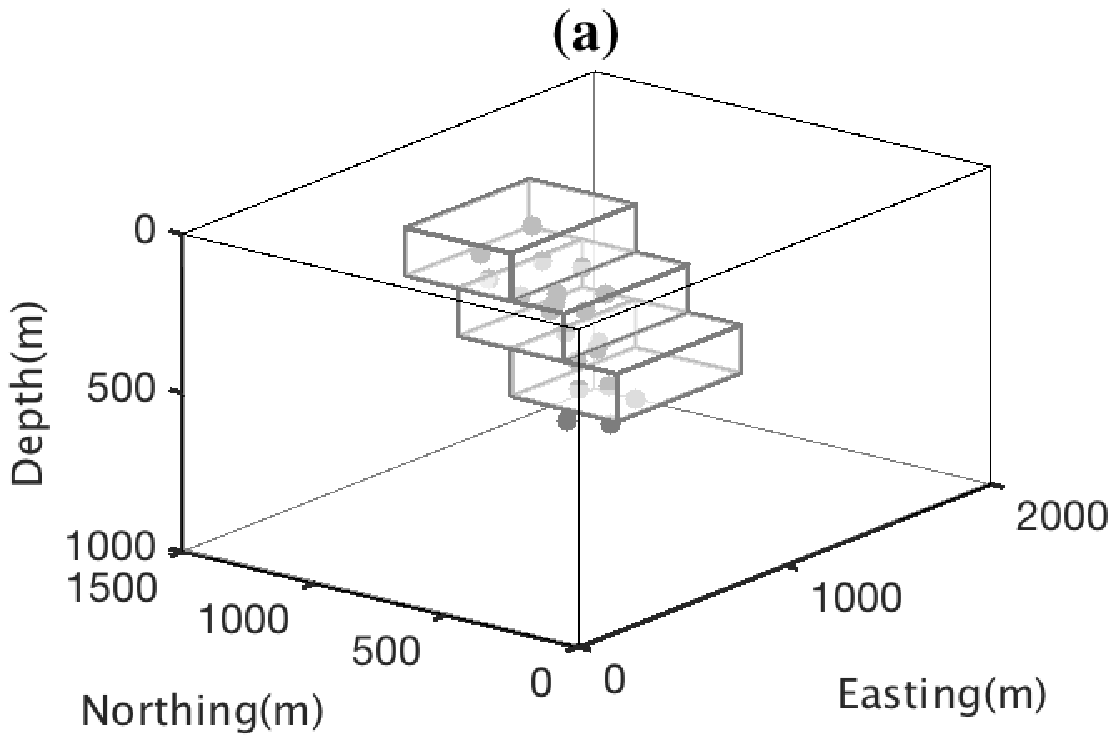}}
\subfloat{\label{fig5b}\includegraphics[width=0.25\textwidth]{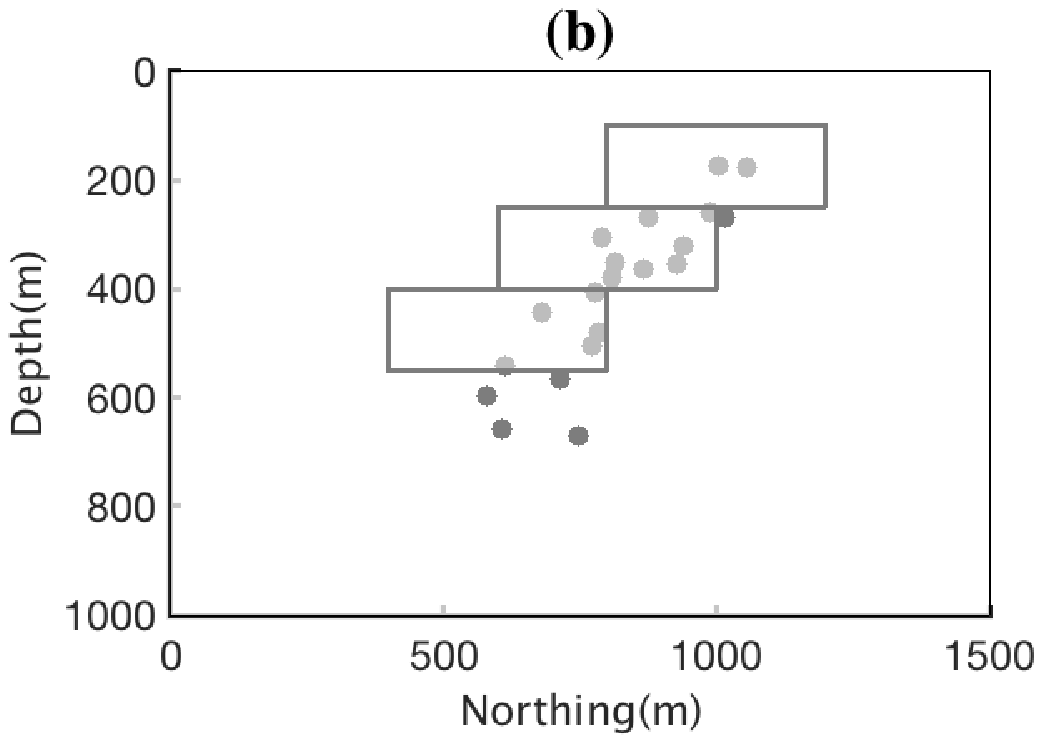}} 
\subfloat{\label{fig5c}\includegraphics[width=0.25\textwidth]{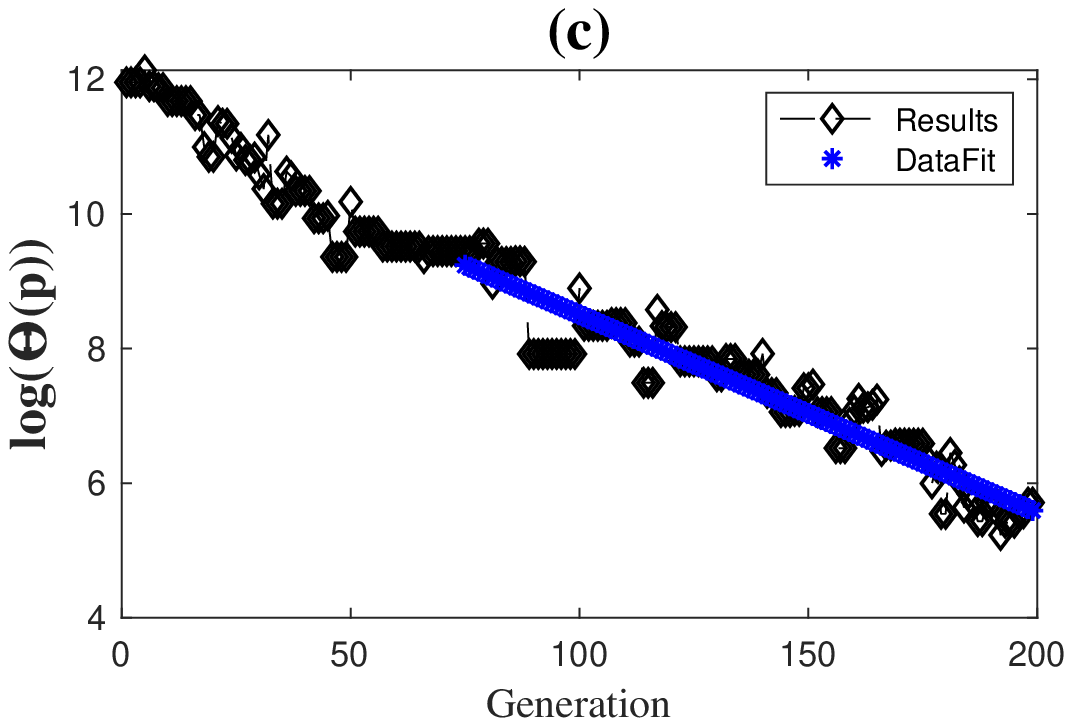}}
\subfloat{\label{fig5d}\includegraphics[width=0.25\textwidth]{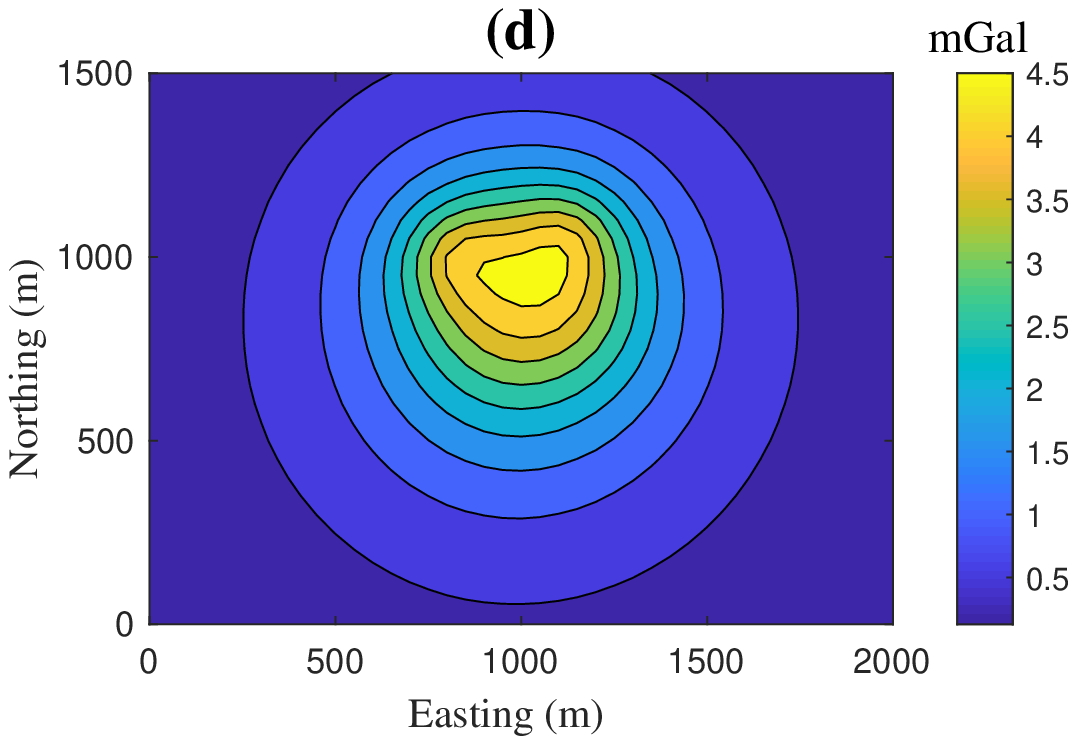}}
\caption{ Inversion results using Algorithm~\ref{algorithm1} with regularization parameter $.1$, $\Kmax=200$ and $M=20$. (a) A perspective view of the point masses; (b) The cross-sectional view of the point masses; (c) The equidistance function for the best solution at each iteration; (d) The data predicted  by the reconstructed model.}\label{fig5}
\end{figure}

\begin{figure}[H]
\subfloat{\label{fig6a}\includegraphics[width=0.25\textwidth]{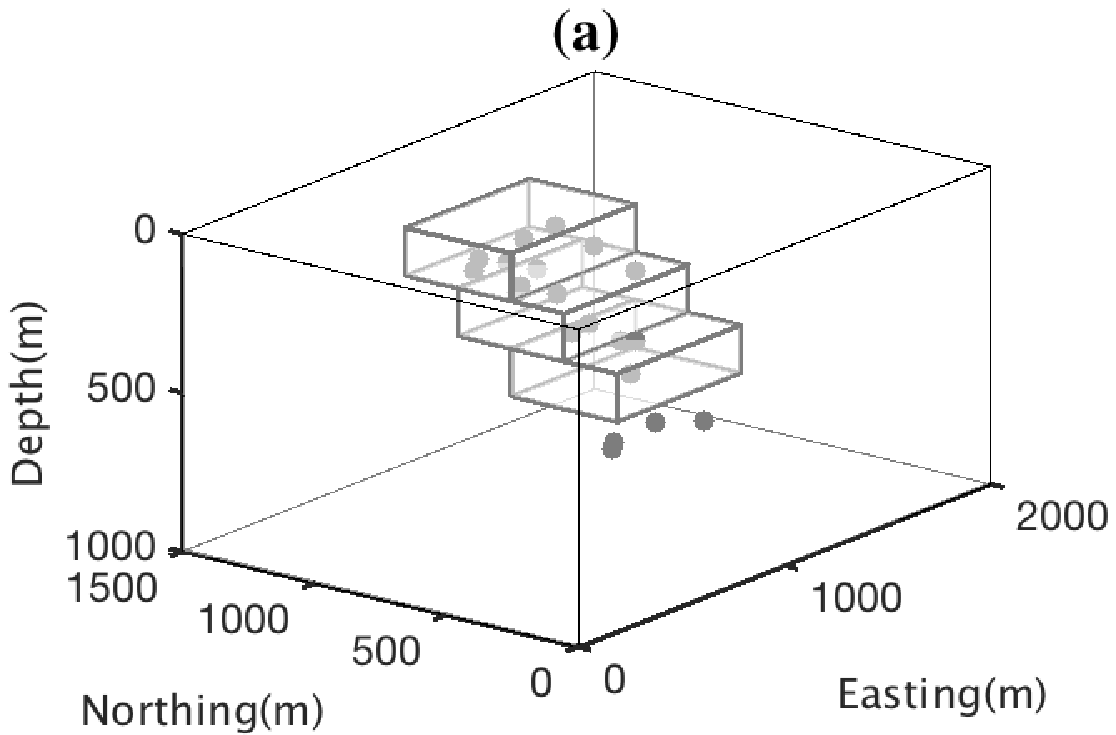}}
\subfloat{\label{fig6b}\includegraphics[width=0.25\textwidth]{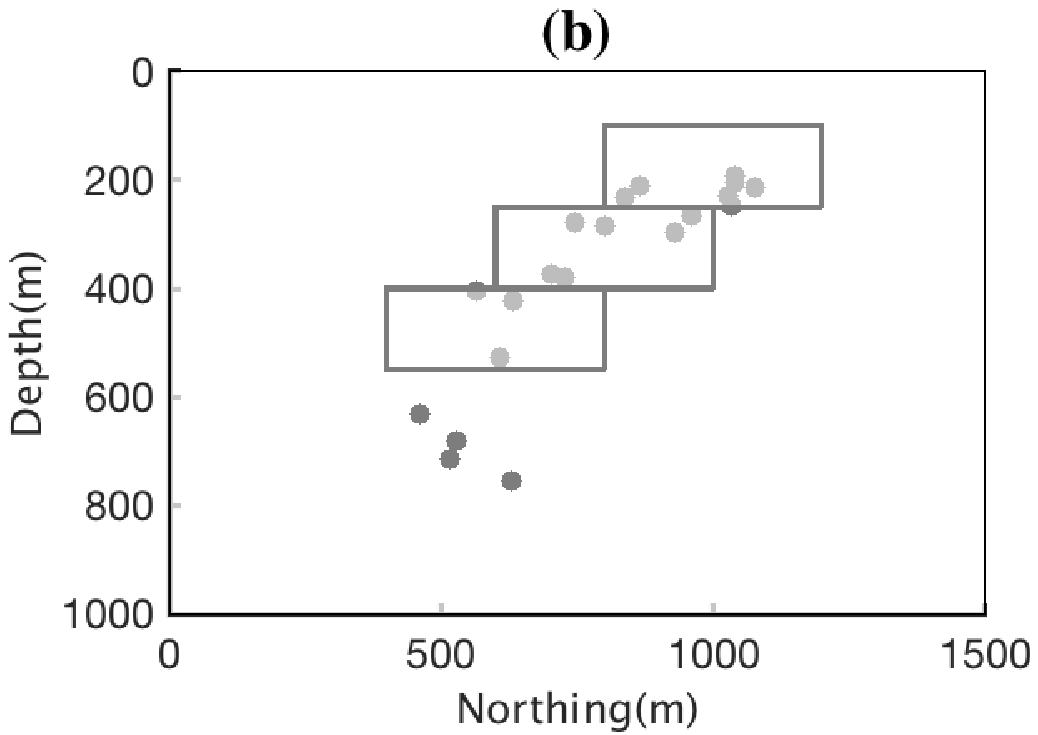}} 
\subfloat{\label{fig6c}\includegraphics[width=0.25\textwidth]{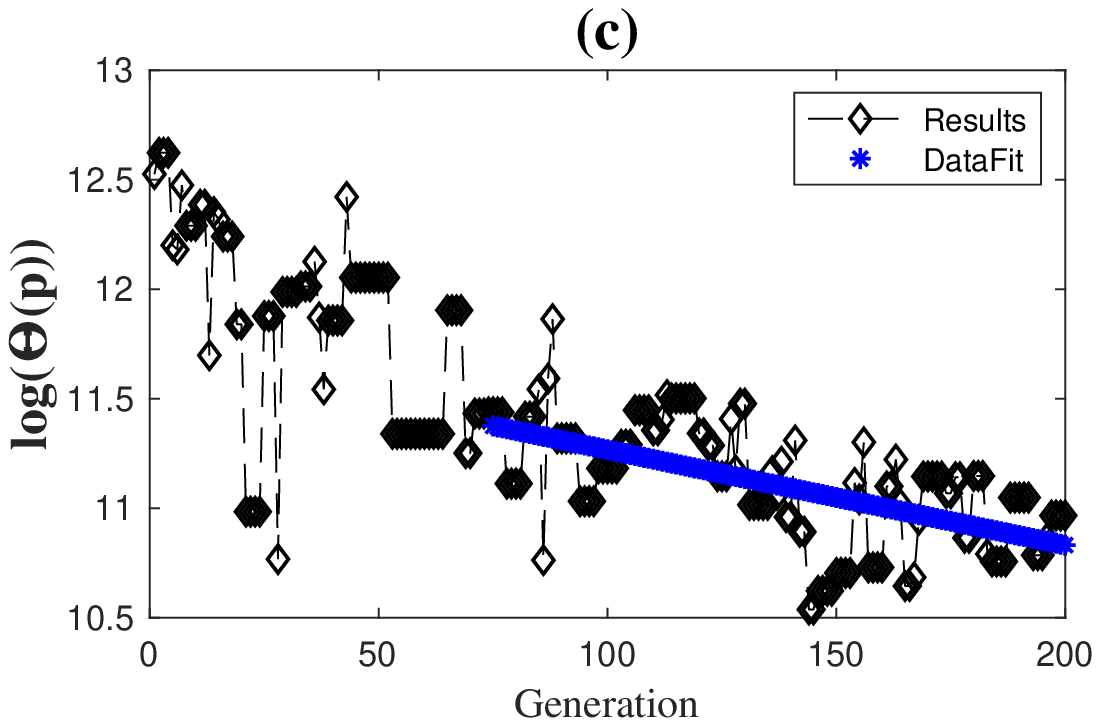}}
\subfloat{\label{fig6d}\includegraphics[width=0.25\textwidth]{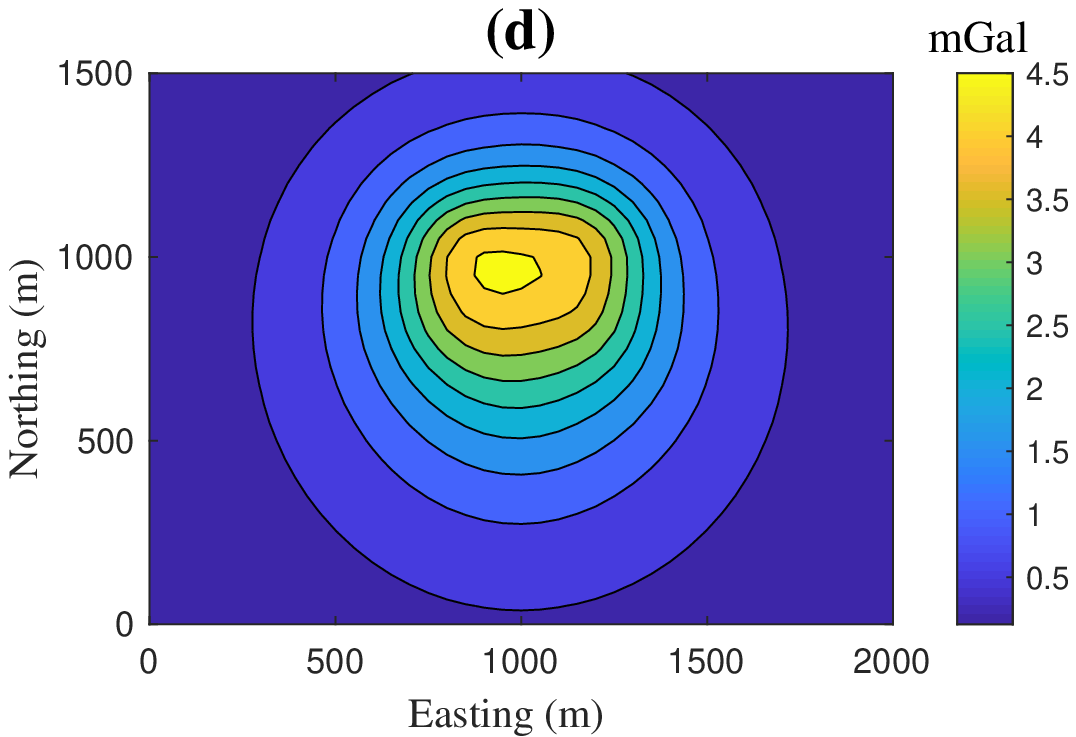}}
\caption{Inversion results using Algorithm~\ref{algorithm1} with regularization parameter $1 \times 10^{-5}$,  $\Kmax=200$ and $M=20$. (a) A perspective view of the point masses; (b) The cross-sectional view of the point masses; (c) The equidistance function for the best solution at each iteration; (d) The data predicted  by the reconstructed model.}\label{fig6}
\end{figure}

\begin{figure}[H]
\subfloat{\label{fig7a}\includegraphics[width=0.25\textwidth]{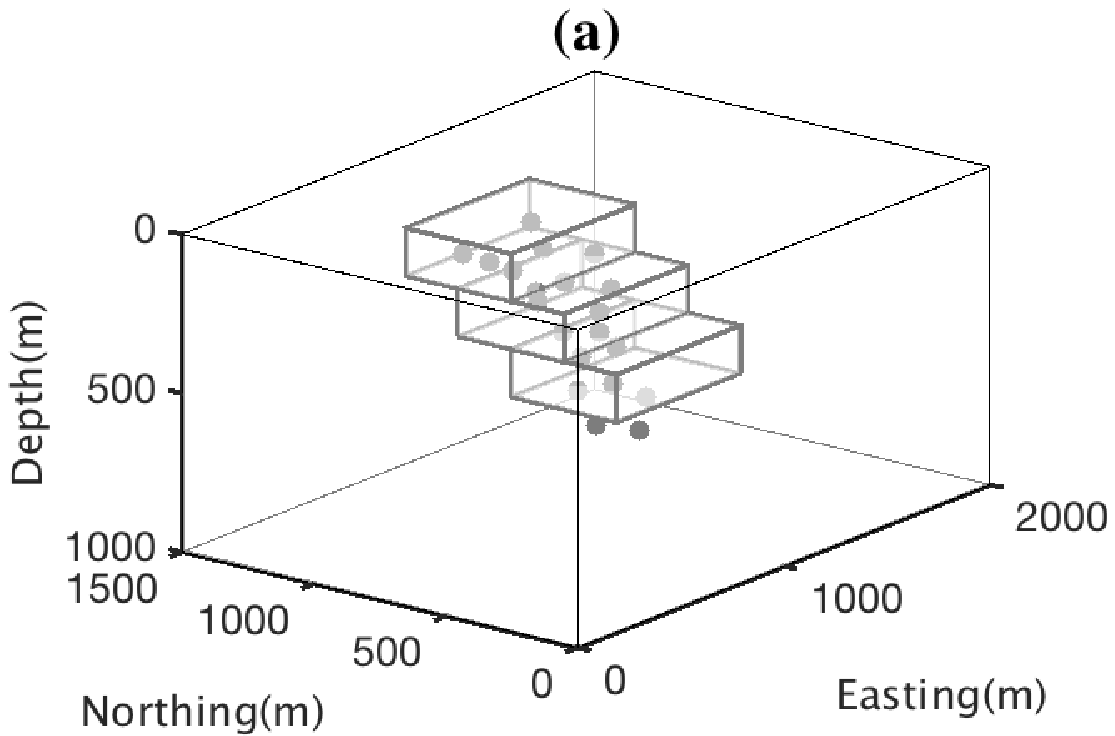}}
\subfloat{\label{fig7b}\includegraphics[width=0.25\textwidth]{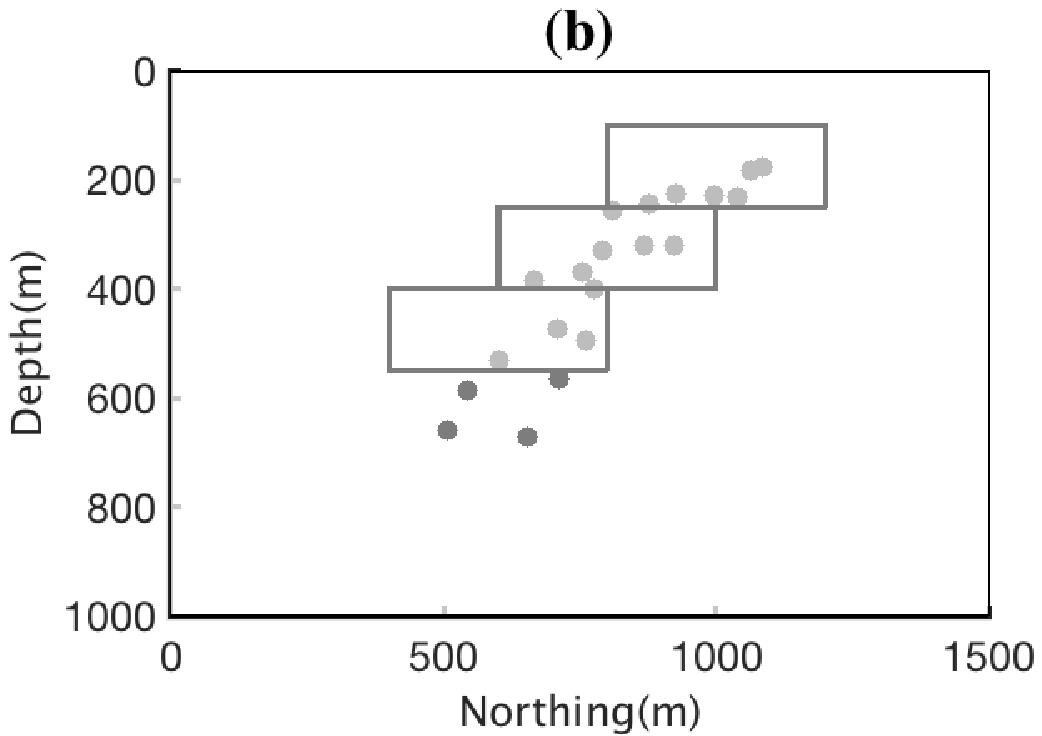}} 
\subfloat{\label{fig7c}\includegraphics[width=0.25\textwidth]{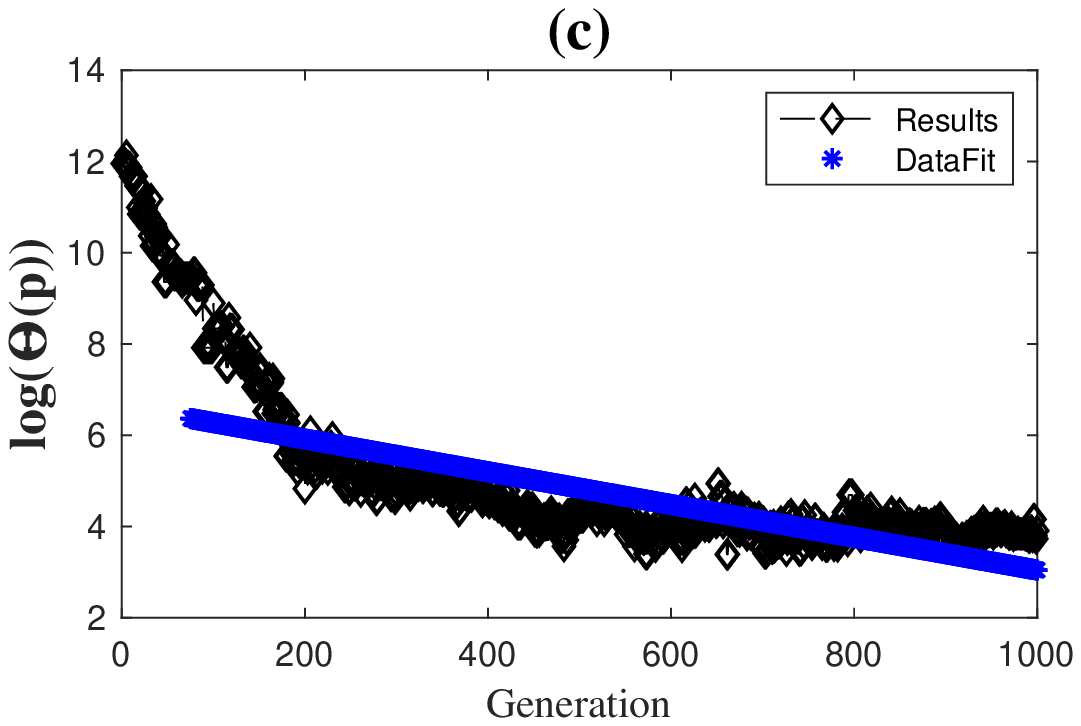}}
\subfloat{\label{fig7d}\includegraphics[width=0.25\textwidth]{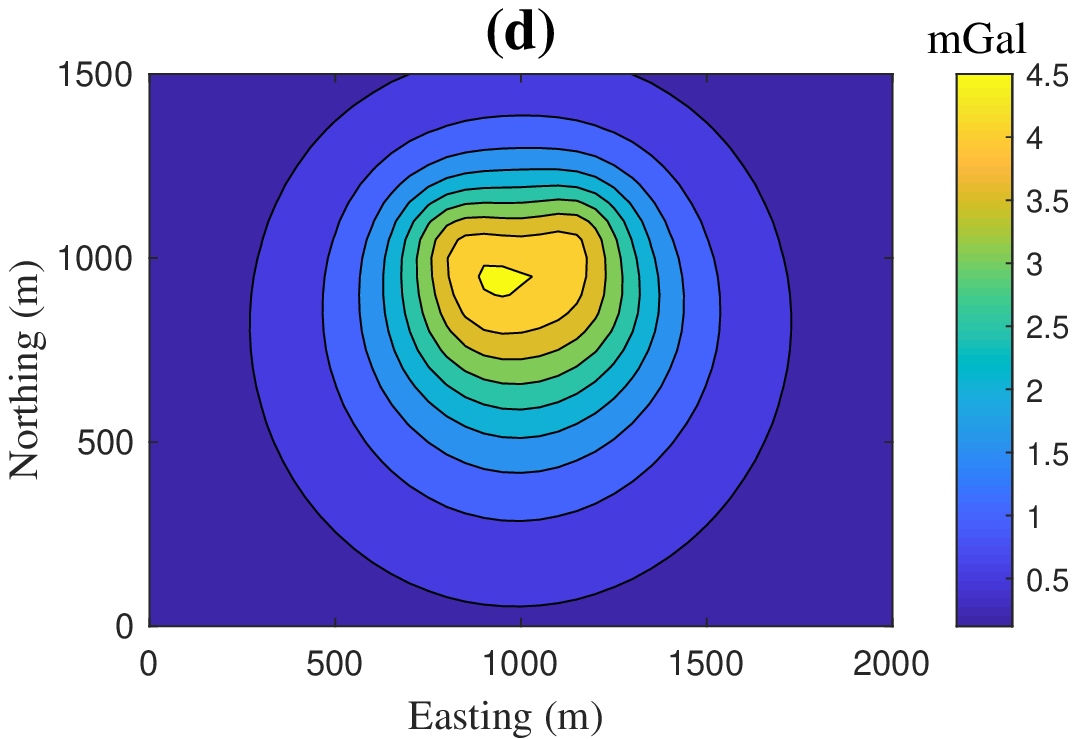}}
\caption{ Inversion results using Algorithm~\ref{algorithm1} with regularization parameter $.1$, $\Kmax=1000$ and $M=20$. (a) A perspective view of the point masses; (b) The cross-sectional view of the point masses; (c) The equidistance function for the best solution at each iteration; (d) The data predicted  by the reconstructed model.}\label{fig7}
\end{figure}

\begin{figure}[H]
\subfloat{\label{fig8a}\includegraphics[width=0.25\textwidth]{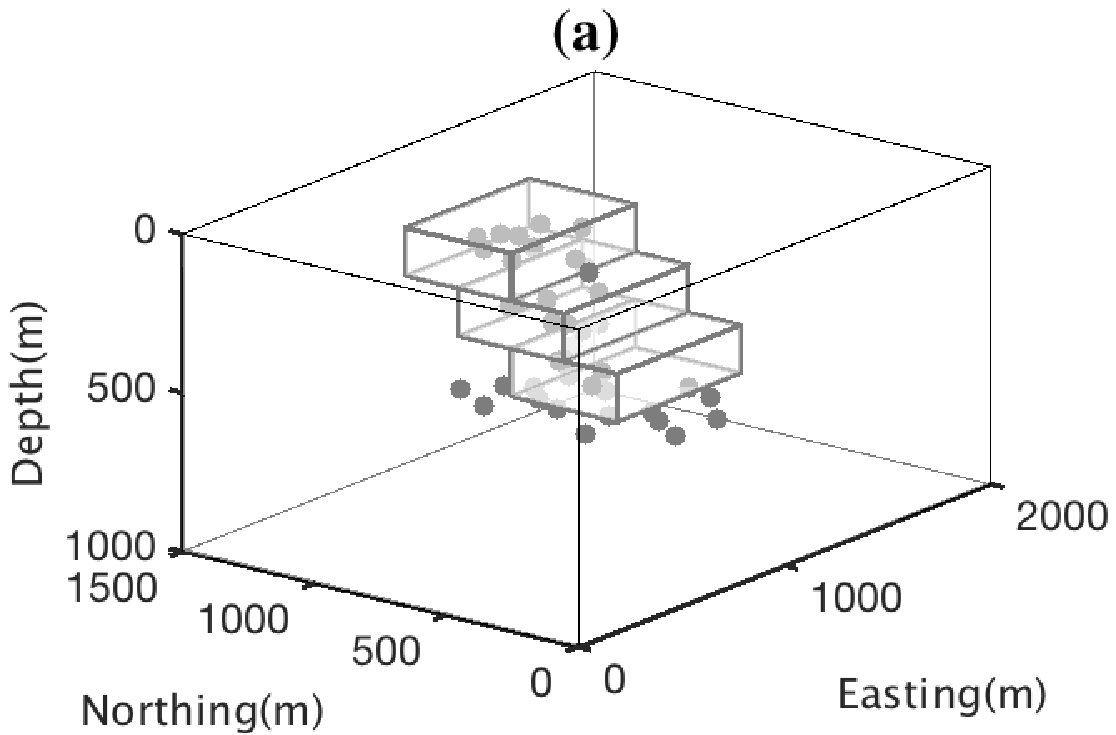}}
\subfloat{\label{fig8b}\includegraphics[width=0.25\textwidth]{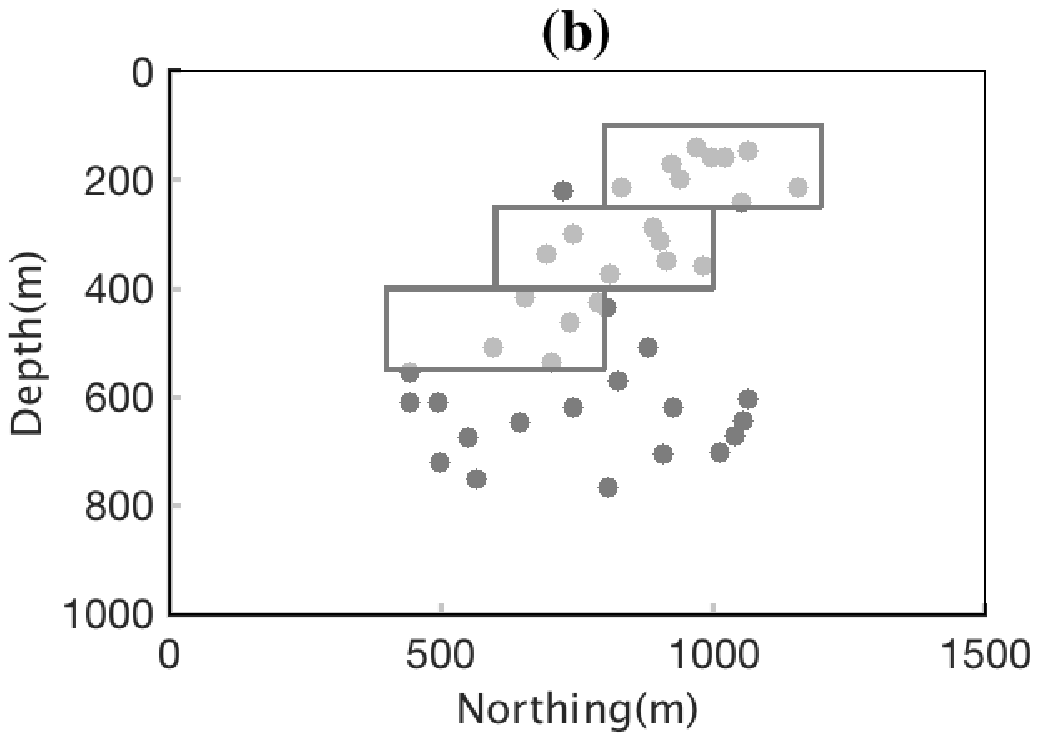}}
\subfloat{\label{fig8c}\includegraphics[width=0.25\textwidth]{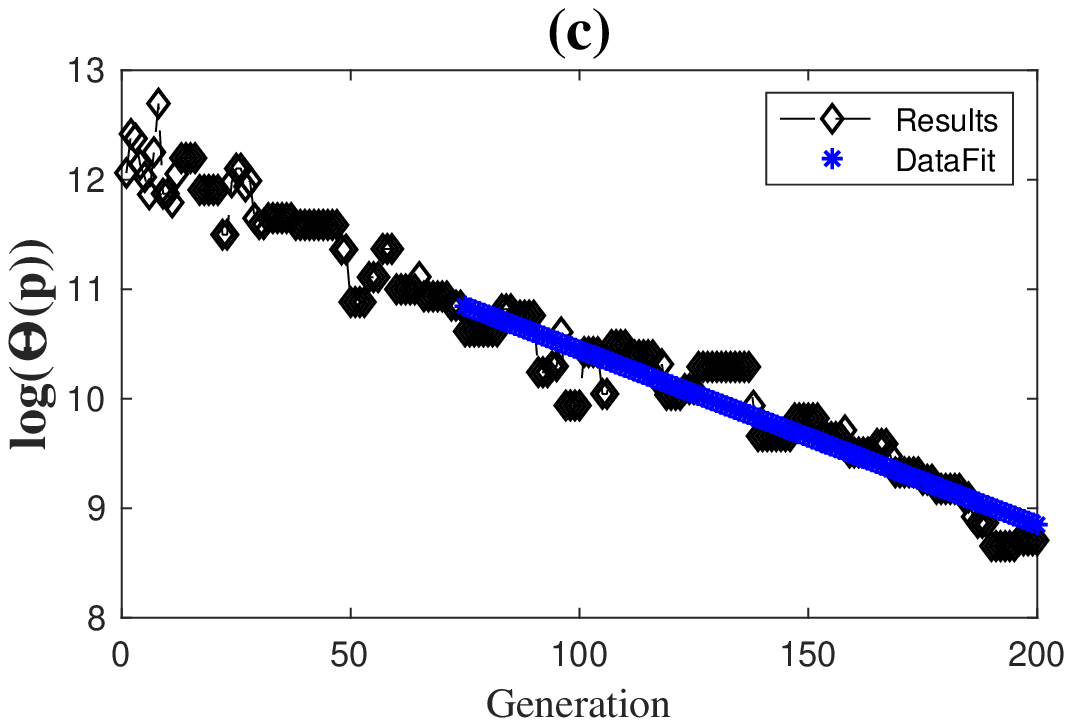}}
\subfloat{\label{fig8d}\includegraphics[width=0.25\textwidth]{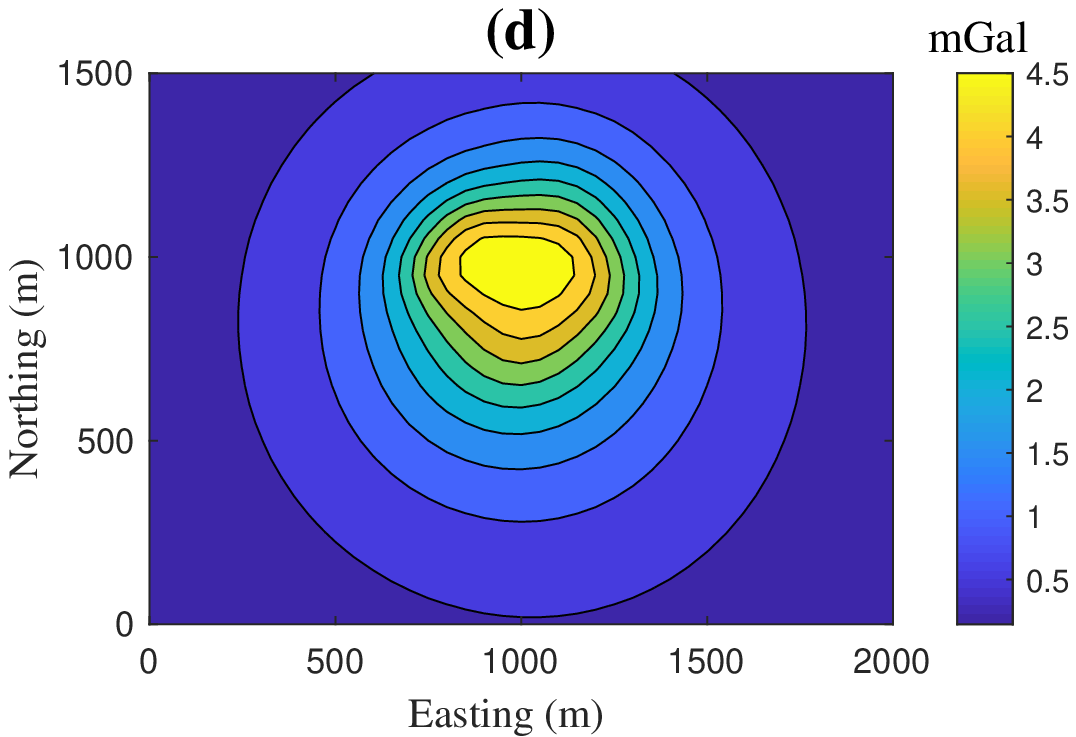}}
\caption{Inversion results using Algorithm~\ref{algorithm1} with regularization parameter $.1$, $\Kmax=200$ and $M=40$. (a) A perspective view of the point masses; (b) The cross-sectional view of the point masses; (c) The equidistance function for the best solution at each iteration; (d) The data predicted  by the reconstructed model.}\label{fig8}
\end{figure}


\subsection{Applying Diagnostics to Determine $\lambda$}\label{diagnostics}
We now discuss an assessment  tool implemented in \texttt{Diagnostic$\_$Results.m} that can be used to analyze the results  based on a regression (linear data fit). This provides a computationally efficient method to identify a $\lambda$ that provides a solution that is neither under or over regularized, without performing the extensive computation required to generate an $L-$curve. First the analysis demonstrates that while both $\Phi$ and $\Gamma$ decay linearly with $k$, so we can use  \eqref{eq.ls}, $\Theta$ decays proportionally to $A\exp(-k)$, and thus regression is applied for  $\log(\Theta(k)) = a k +b$, where $A=\exp(b)$.  

Table~\ref{tab4} gives the results for a model simulation obtained for all the parameters as given in Tables~\ref{tab1} and \ref{tab2} for the simulation illustrated in  Fig.~\ref{fig3b} an inversion with $20$ mass points, maximum iteration $\Kmax=200$ and the noted range of $\lambda_i$. From the results in Table~\ref{tab4} it is evident that the convergence behavior of $\Theta$ is stable for large $\lambda$;  $R^2$ is close to $1$ but $\Phi$ is large relative to the noise level and the mass estimation is not stable, the mass may be underestimated. Further, for large $\lambda$ the solution terminates with small $\Theta$.   The $R^2$ value  eventually decreases as $\lambda$ decreases before increasing again at the choice of $\Phi$ which is closest to the noise estimate.  These results suggest that an acceptable solution will be obtained for $\lambda$ ranging from about $0.1 $ to $.025$. We illustrate the  resulting mass point distributions for $\lambda=10$, $.5$, and $.025$ in Fig.~\ref{fig9}, demonstrating that the analysis is relevant.  There are also links to \href{https://math.la.asu.edu/~rosie/research/gravity/htmlruns/SimulatedData}{simulated data sets} giving several analyses of data for multiple choices of $\lambda$, $M$ and noise levels in the accompanying webpage. 
 \begin{table} [H]
\scriptsize
\caption{The results of the inversion of the model for the given selections of $\lambda$, $M=20$ and $\Kmax=200$. The total time for the inversions reported in the table is $1452.2$ seconds, or approximately $24$ minutes. }
\centering
\begin{tabular}{|c|c|c|c|c|c|c|c|c|c|c|}
\hline
  $\lambda_i$ & $k$ & mass & $\Theta(k)$ & $\Phi(k)$&$\Phi(k)/(N+\sqrt{2N}) $ &$R^2$ \\ \hline
   $  100$ & $  200$ & $  112.6e+9$ & $  17.486$ & $  54315$ &$41.1 $ & $  0.88  $\\ \hline
    $    10$ & $  200$ & $  83.8e+9$ & $  9.86$ & $  75172$ &$56.9 $ & $  0.93 $\\ \hline
    $     1$ & $  200$ & $  141.4e+9$ & $   70.517$ & $ 11427$ & $8.65 $ &$  0.91$\\ \hline
 $     0.5$ & $  200$ & $   120.0e+9$ & $   109.03$ & $  4746.8$ &$3.59 $ & $  0.96$\\ \hline
 $     0.25$ & $  200$ & $  127.5e+9$ & $   202.24$ & $  3646.8$ & $2.76 $ &$  0.95$\\ \hline
$       0.1$ & $  200$ & $  119.7e+9$ & $   381.8$ & $  2129.8$ & $1.61$ &$  0.95 $\\ \hline
 $     0.05$ & $  200$ & $  118.0e+9$ & $   373.48$ & $ 1898.4$ &$1.44$ & $  0.86$\\ \hline
$     0.025$ & $  200$ & $  122.5e+9$ & $   1262$ & $  2492.2$ & $1.89 $ &$  0.94 $\\ \hline
 $     0.01$ & $  200$ & $  116.9e+9$ & $3858.4$ & $1608.6  $ & $1.22 $ &$0.88$\\ \hline
 $    0.001$ & $  200$ & $  114.2e+9$ & $    14654$ & $  1484.0$ &$1.12 $ & $   0.86$ \\ \hline
 $   0.0001$ & $  200$ & $  118.0e+9$ & $    51524$ & $ 1512.3$ & $1.14 $ &$   0.01 $\\ \hline
 $  0.00001$ & $   200$ & $  117.9e+9$ & $   300060$ & $  1543.2$ & $1.17 $ &$  0.38$\\ \hline
\end{tabular}
\label{tab4}
\end{table}
\begin{figure}[H]
\begin{center}
\subfloat{\label{fig9a}\includegraphics[width=0.3 \textwidth]{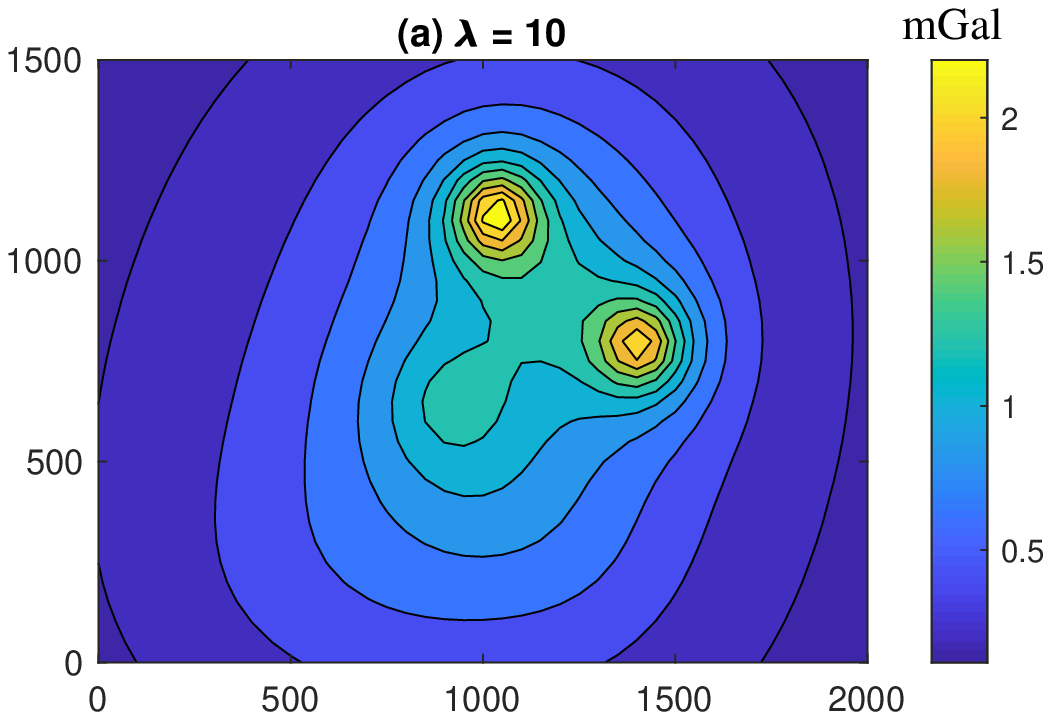}}
\subfloat{\label{fig9b}\includegraphics[width=0.3 \textwidth]{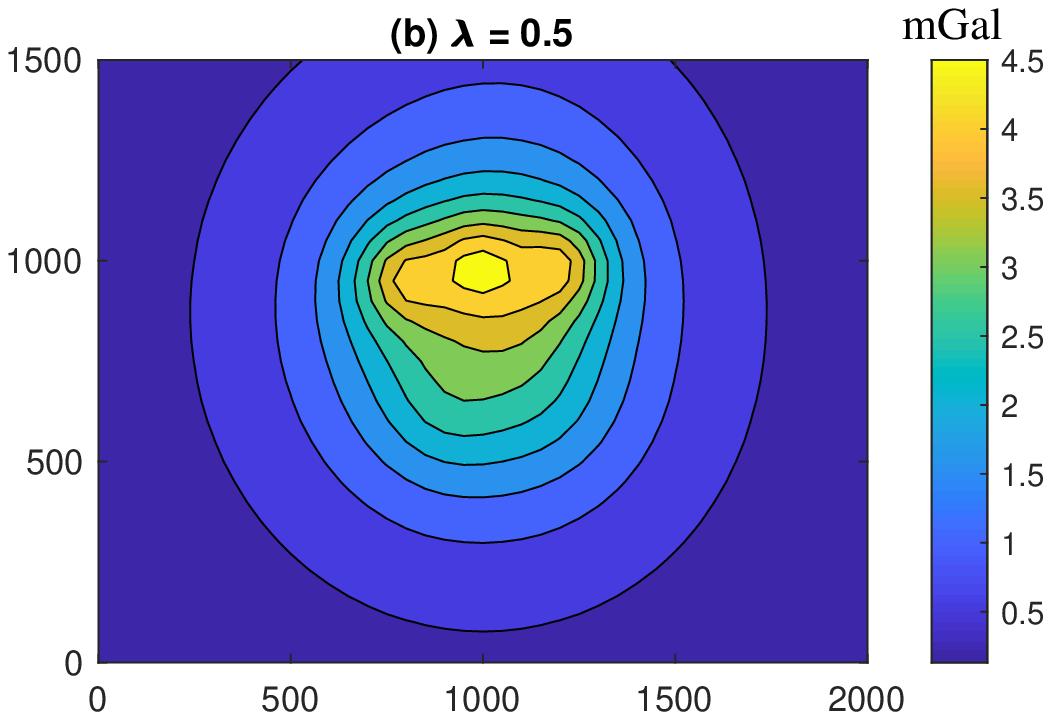}}
\subfloat{\label{fig9c}\includegraphics[width=0.3 \textwidth]{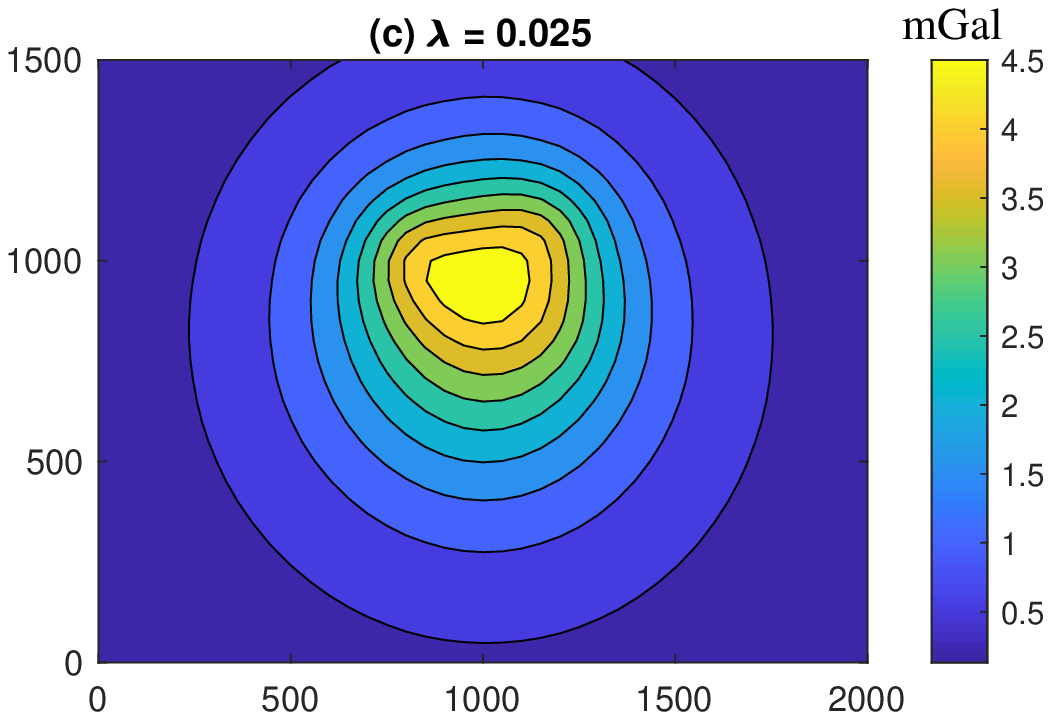}}\\
\subfloat{\label{fig9d}\includegraphics[width=0.3 \textwidth]{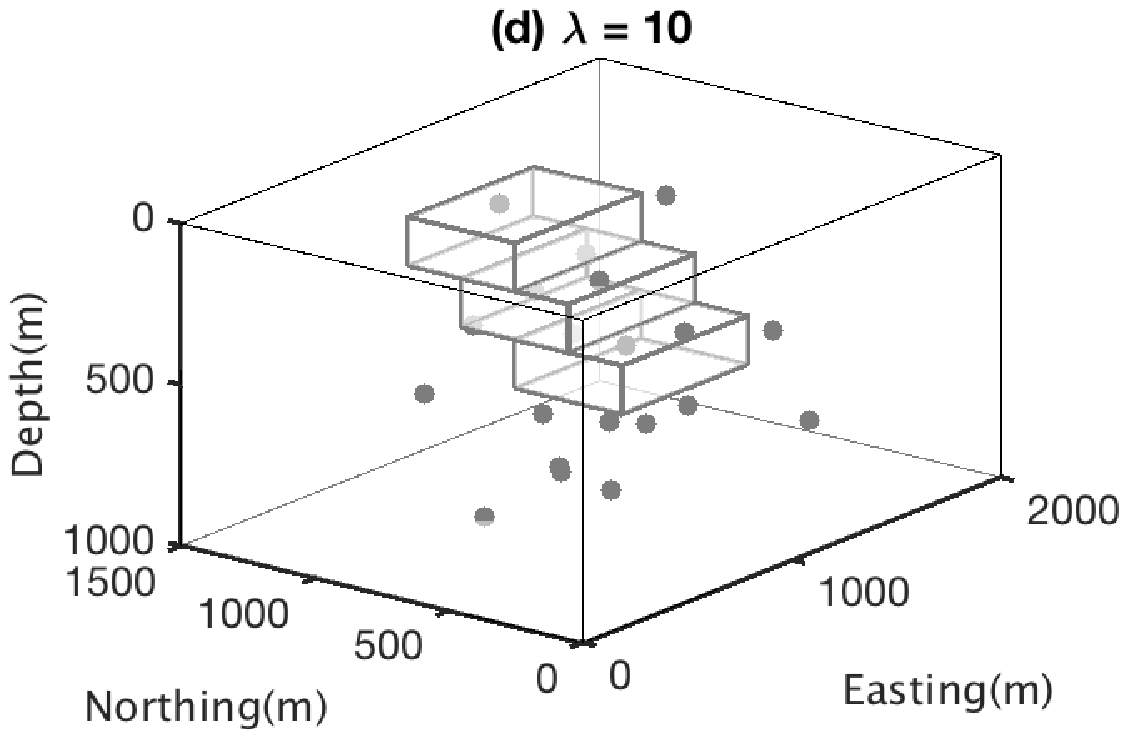}}
\subfloat{\label{fig9e}\includegraphics[width=0.3 \textwidth]{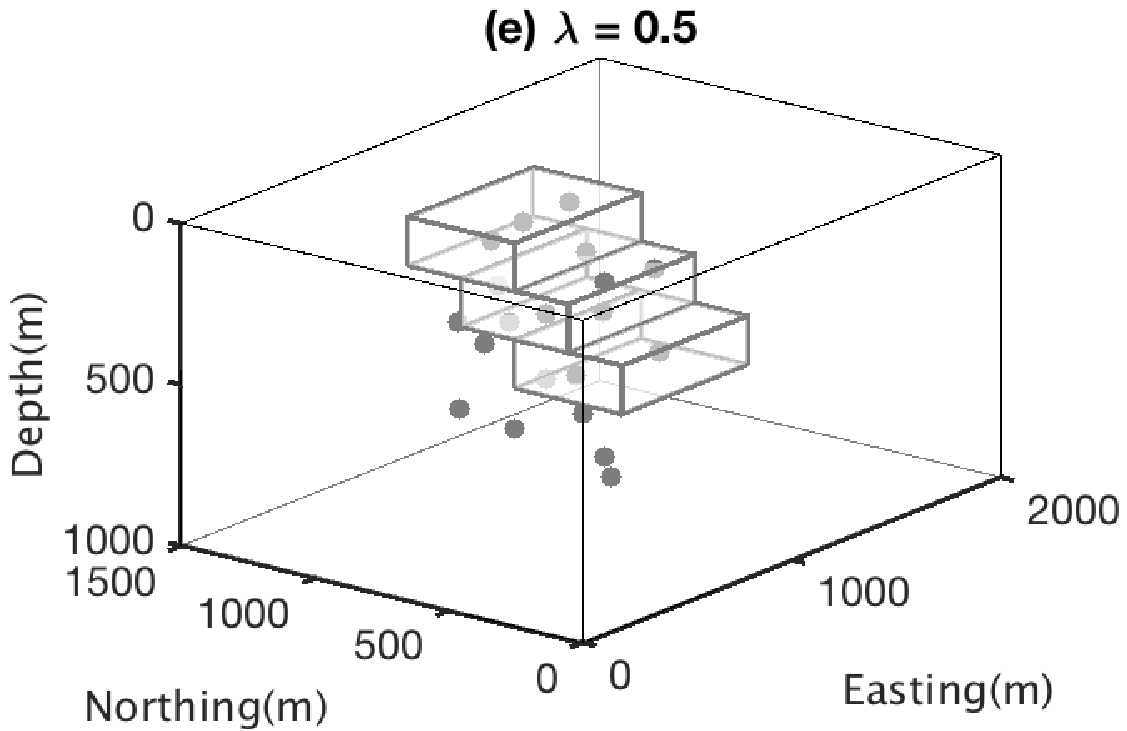}}
\subfloat{\label{fig9f}\includegraphics[width=0.3 \textwidth]{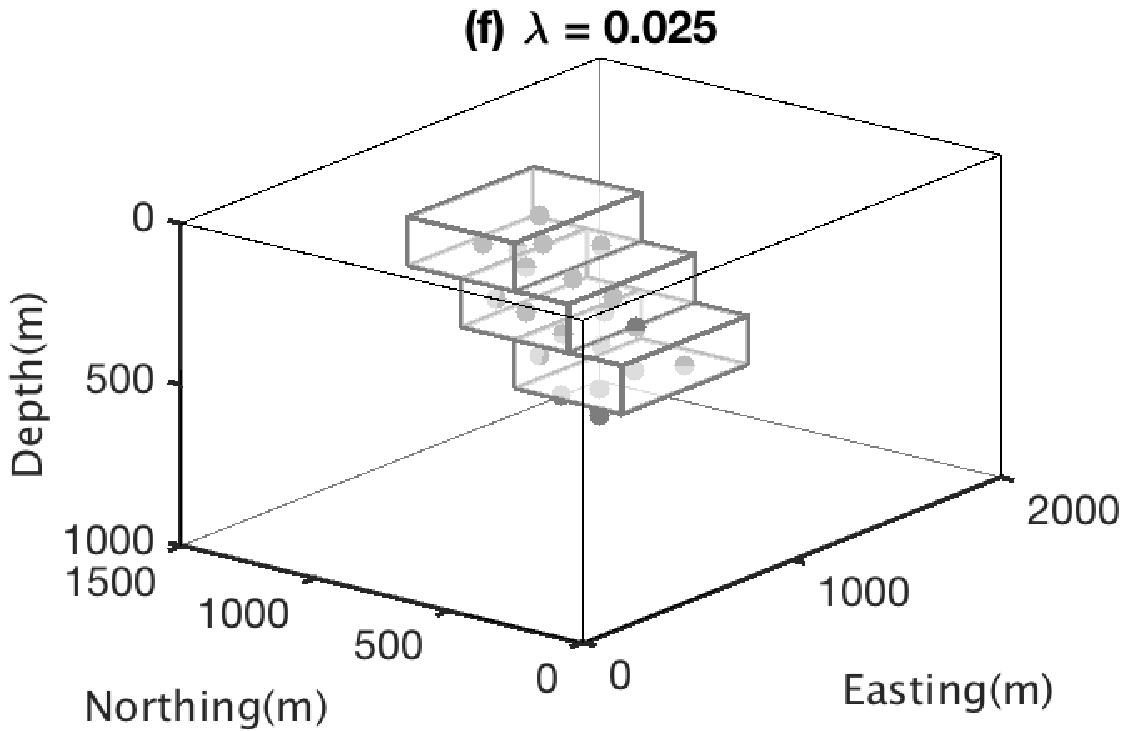}}\\
\subfloat{\label{fig9g}\includegraphics[width=0.3 \textwidth]{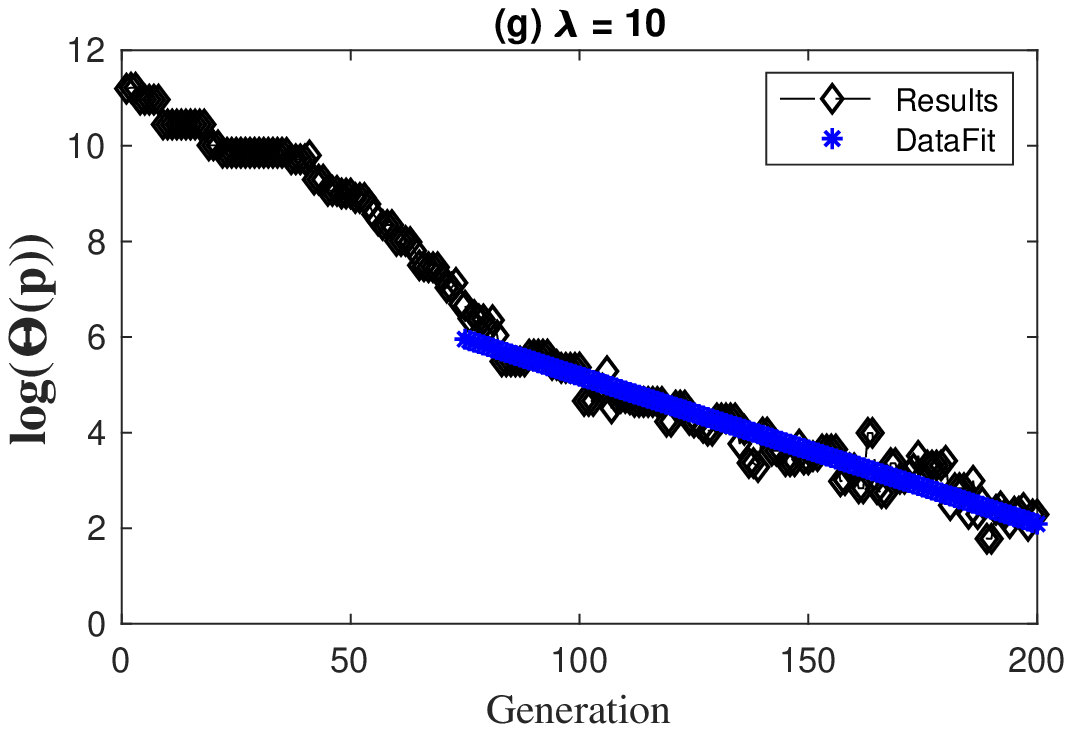}}
\subfloat{\label{fig9h}\includegraphics[width=0.3 \textwidth]{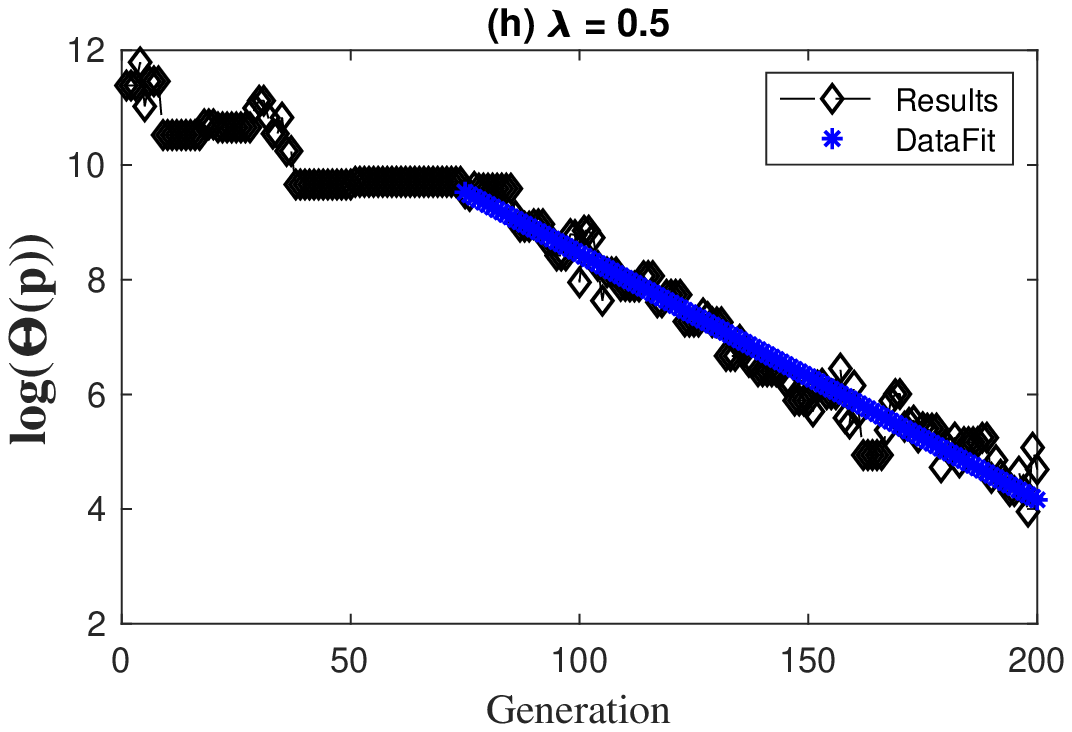}}
\subfloat{\label{fig9i}\includegraphics[width=0.3 \textwidth]{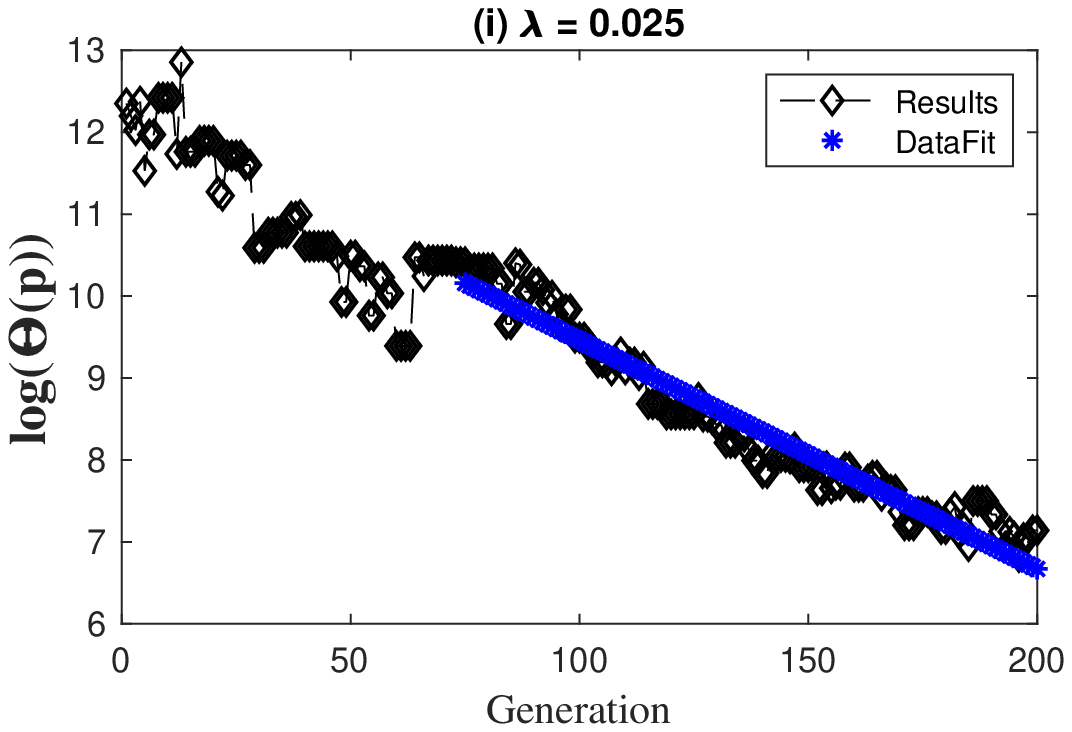}}
\end{center}
\caption{The mass point distributions, the predicted anomalies and $\Theta$ with the indicated regression line (data fit)  for the solutions chosen according to the data in Table~\ref{tab4} for the model illustrated in Fig.~\ref{fig3b}. }\label{fig9}
\end{figure}

\subsection{Real data}\label{real}
To illustrate the relevance of the approach for a practical case we applied software to reconstruct the well-known Mobrun ore body, northeast of Noranda, Quebec, Canada, Fig.~\ref{fig10}. The anomaly pattern is associated with a massive body of base metal sulphide (mainly pyrite) which has displaced volcanic rocks of middle Precambrian age \citep{GrWe:65}. We carefully digitized the data from figure $10.1$ in \cite{GrWe:65}, and re-gridded onto a regular grid of $ 37 \times 31= 1147$ data in east and north directions respectively, with grid spacing $20$~m. We approximate the error distribution with $\sigma_i=( 0.03 (\bfdo)_i + 0.004 \| \bfdo \|_2)$. \cite{GrWe:65} interpreted the body to be about $305$~m in length, slightly more than $30$~m in maximum width and having a maximum depth of
$183$~m. Furthermore, they estimated the total mass of the body to be $2.56 e9$~$ kg$.  The parameters of Algorithm~\ref{algorithm1}  for the inversion  are detailed in Table~\ref{tab5}. 

\begin{figure}[H]
\centering
\subfloat{\includegraphics[width=0.5\textwidth]{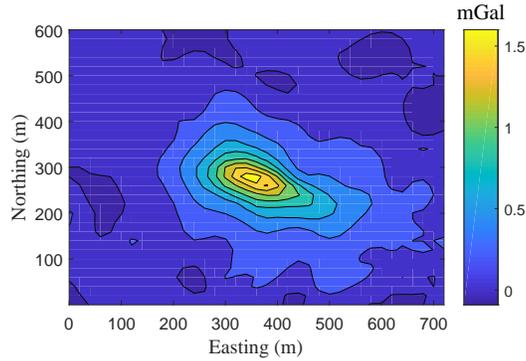}}
\caption{Residual anomaly of Mobrun ore body, Noranda, Quebec, Canada.}\label{fig10}
\end{figure}

\begin{table} [H]
\scriptsize
\caption{Parameters used in Algorithm~\ref{algorithm1} to perform inversion on data of Fig.~\ref{fig10}. Coordinates are given in meters, $m$, and mass in kilograms, $kg$.}
\centering
\begin{tabular}{|c|c|c|c|c|c|c|c|c|c|c|}
\hline
$M$ & $\Kmax$ & $\mathrm{noq}$ & $x_{{\mathrm{min}}}$ & $x_{{\mathrm{max}}} $& $y_{{\mathrm{min}}} $& $y_{{\mathrm{max}}} $& $z_{{\mathrm{min}}} $& $z_{{\mathrm{max}}}$ & $m_{t_{\mathrm{min}}} $& $m_{t_{\mathrm{max}}} $ \\
\hline
 $20$& $200$ & $100$& $150$& $650$& $50$ & $500$& $10$& $300$ & $2.2e9$ & {$3.2e9$} \\ 
\hline 
\end{tabular}
\label{tab5}
\end{table}

We performed the inversion with several fixed values of $\lambda$ and here show the diagnostic results obtained using the selection $\lambda=[10, .25, .001]$ in Table~\ref{tab6}. The resulting mass point distributions and anomalies support the selection of $\lambda=.25$ for the acceptable result.   

\begin{figure}[H]
\begin{center}
\subfloat{\label{fig11a}\includegraphics[width=0.3 \textwidth]{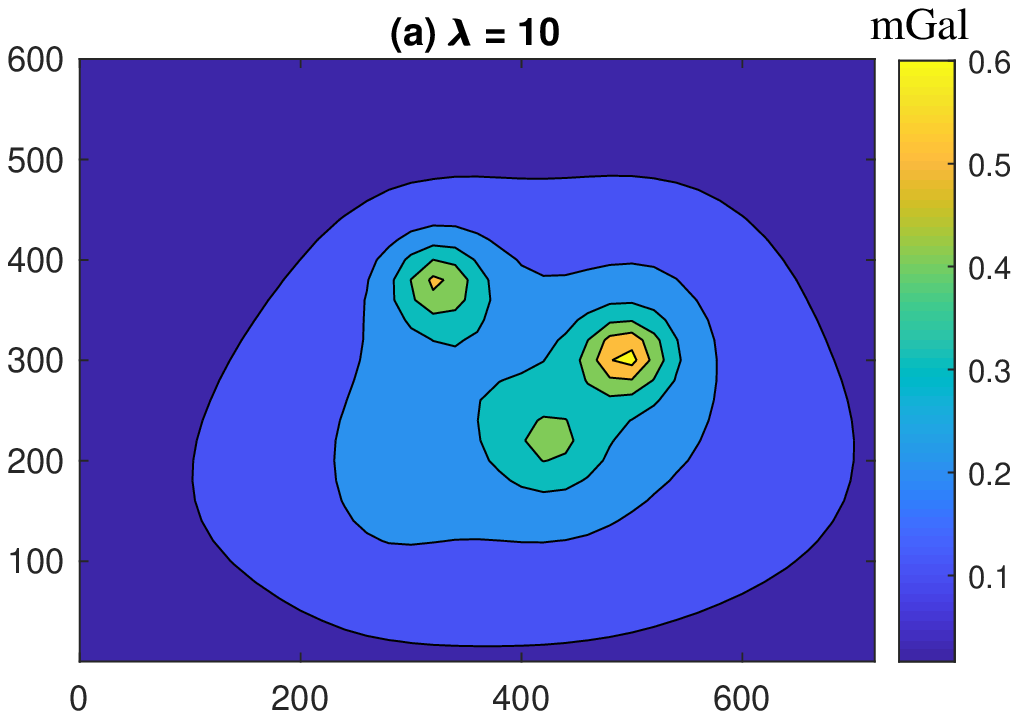}}
\subfloat{\label{fig11b}\includegraphics[width=0.3 \textwidth]{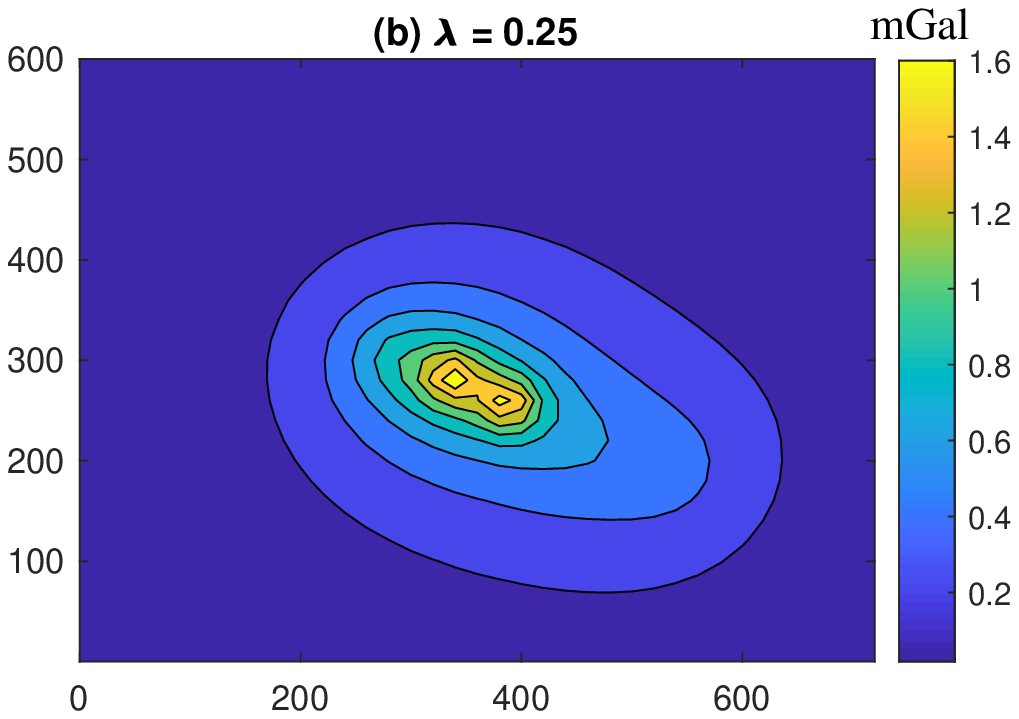}}
\subfloat{\label{fig11c}\includegraphics[width=0.3 \textwidth]{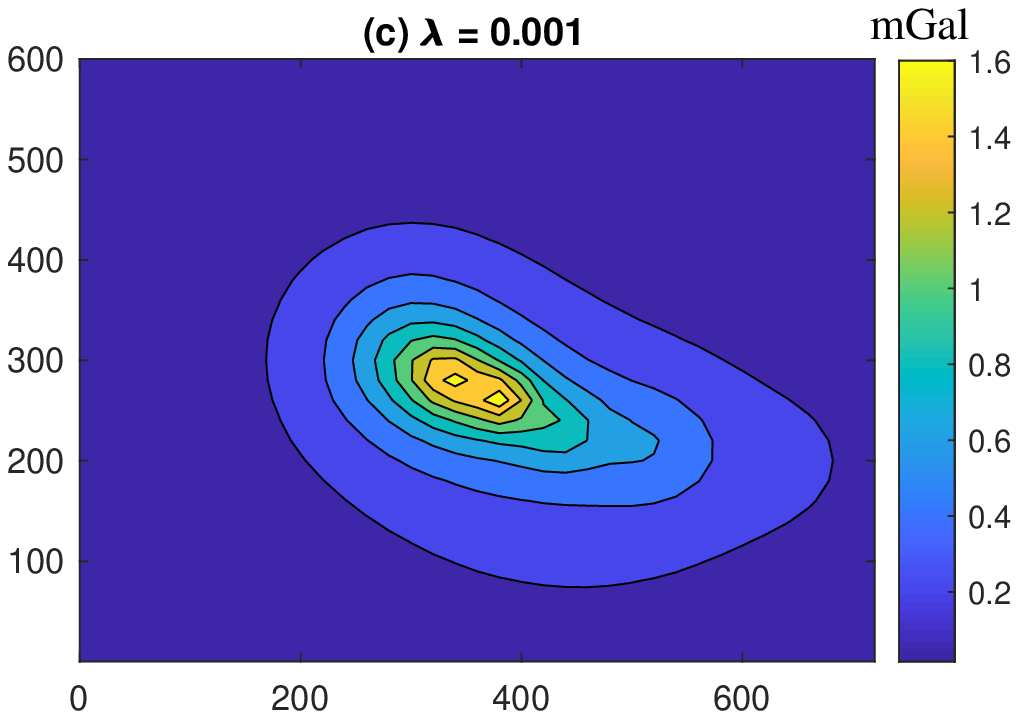}}\\
\subfloat{\label{fig11d}\includegraphics[width=0.3 \textwidth]{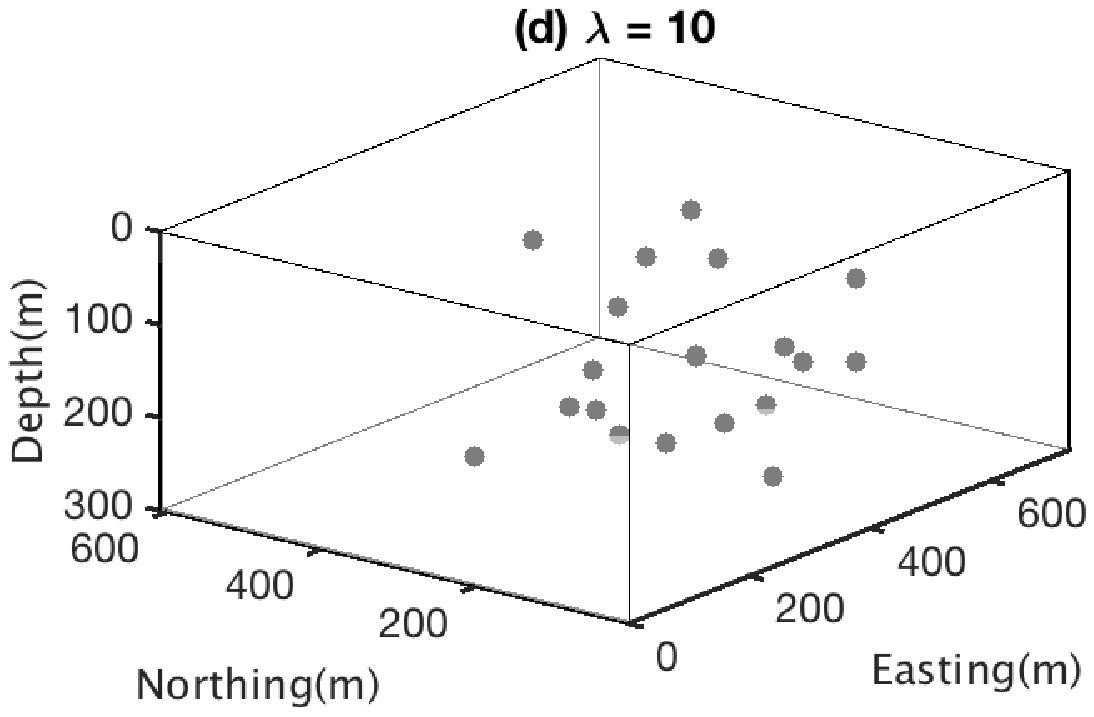}}
\subfloat{\label{fig11e}\includegraphics[width=0.3 \textwidth]{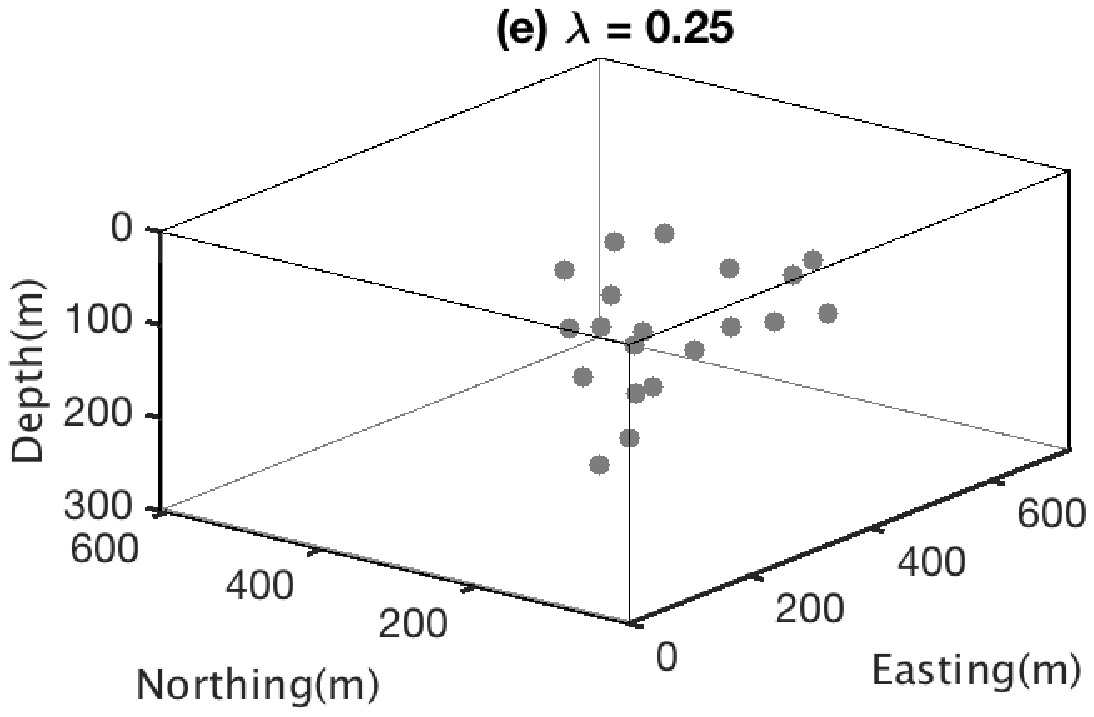}}
\subfloat{\label{fig11f}\includegraphics[width=0.3 \textwidth]{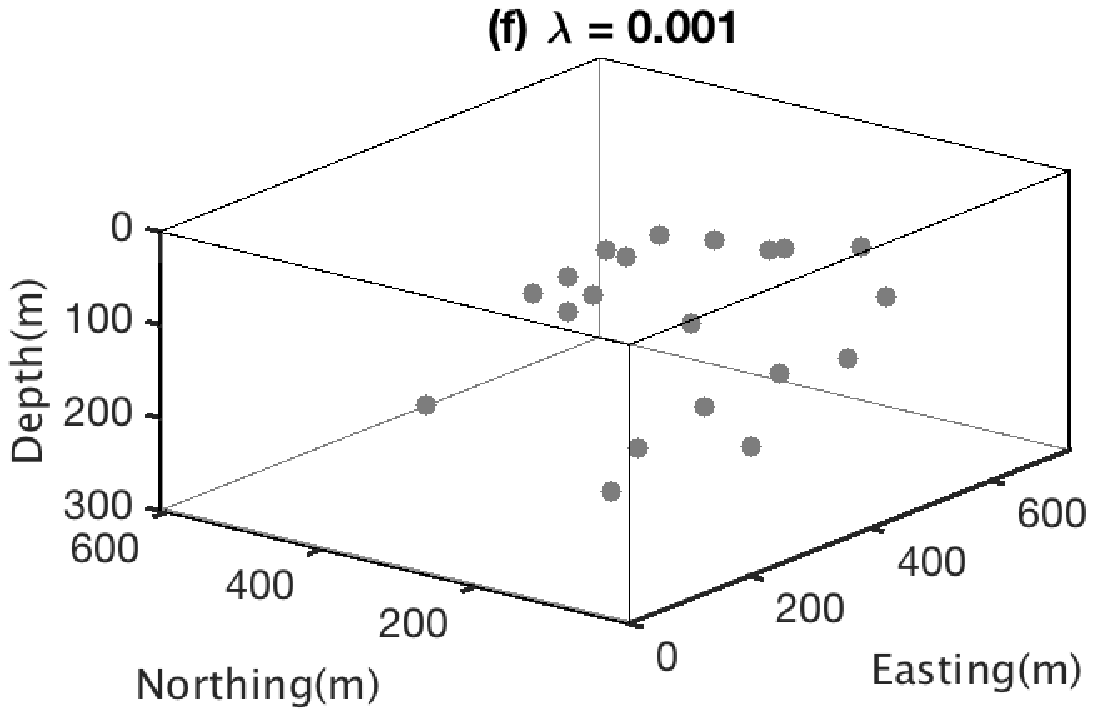}}\\
\subfloat{\label{fig11g}\includegraphics[width=0.3 \textwidth]{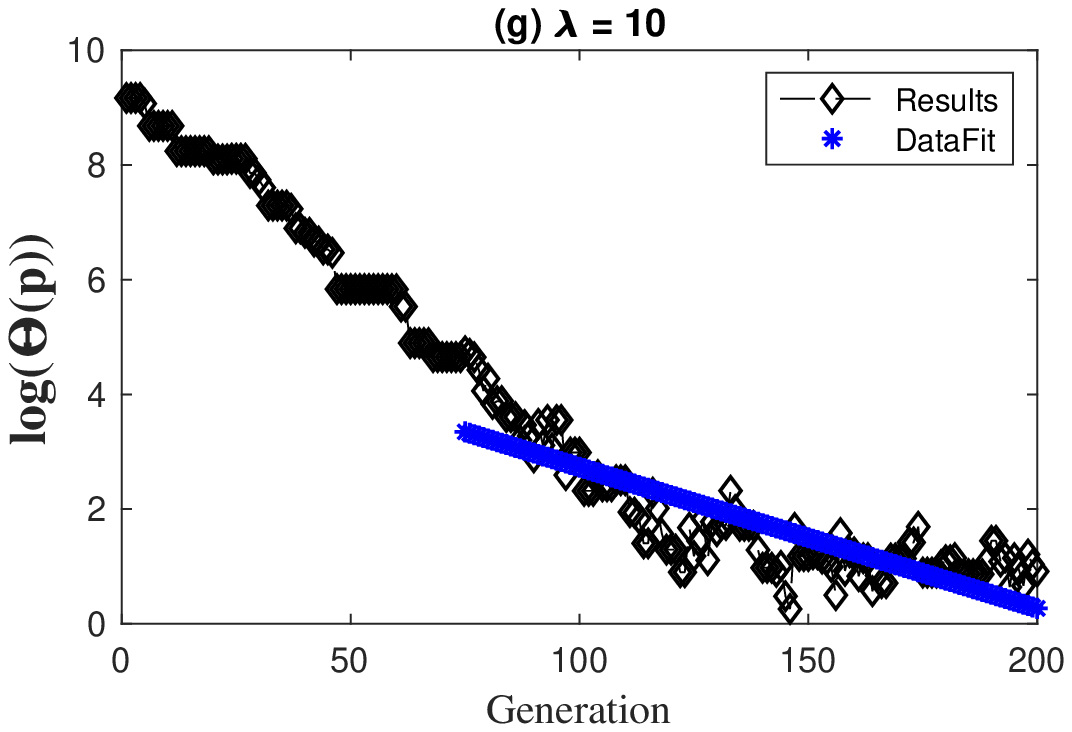}}
\subfloat{\label{fig11h}\includegraphics[width=0.3 \textwidth]{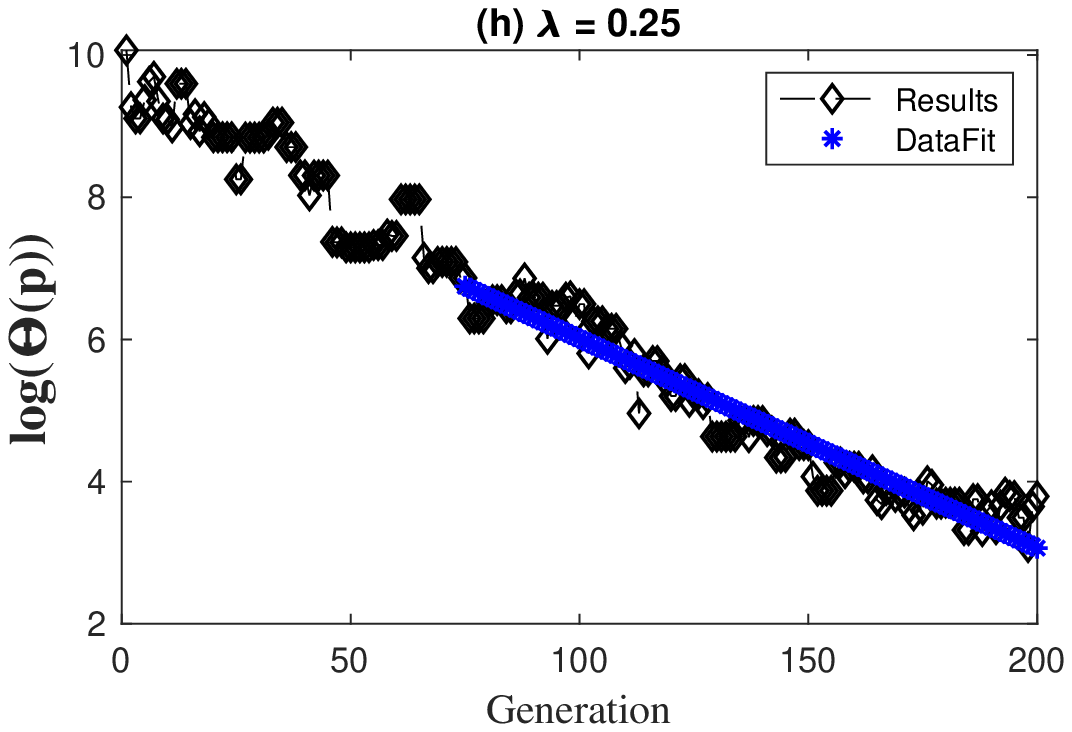}}
\subfloat{\label{fig11i}\includegraphics[width=0.3 \textwidth]{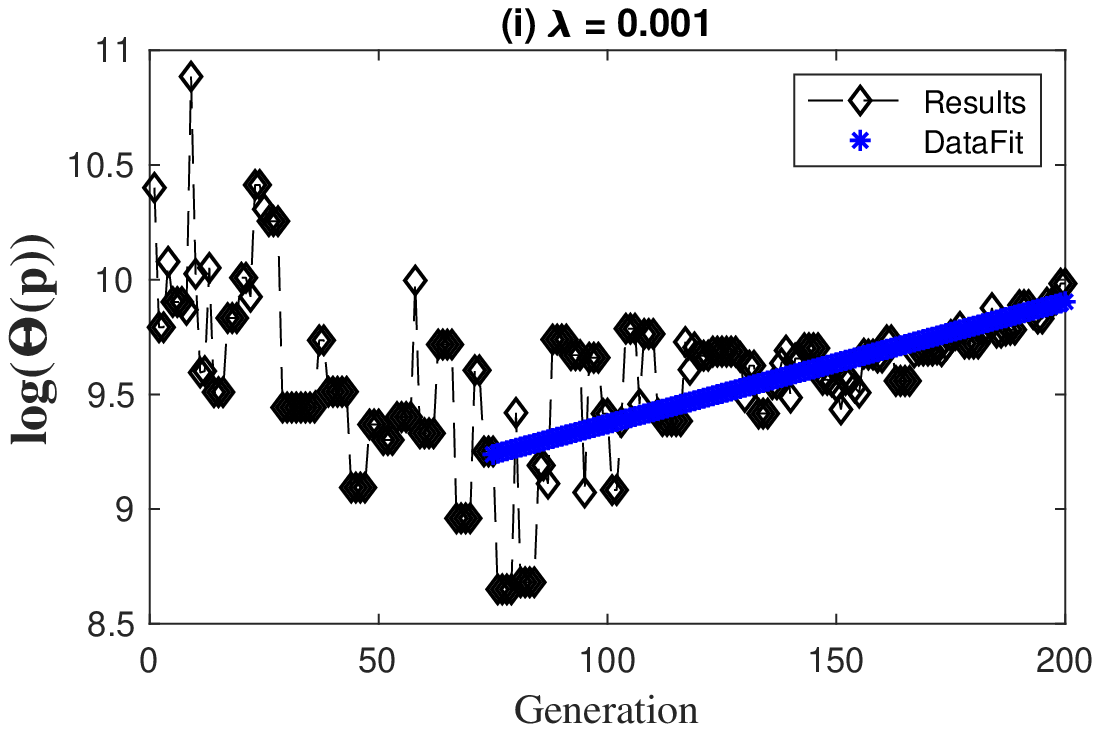}}
\end{center}
\caption{The mass point distributions, predicted anomalies and $\Theta$  with the indicated regression line (data fit)   for the solutions chosen according to the data in Table~\ref{tab5} for the real data  illustrated in Fig.~\ref{fig10}. }\label{fig11}
\end{figure}

 \begin{table} [H]
\scriptsize
\caption{The results of the inversion for real data. The total time for the inversion for all values of $\lambda$ is $509 $ seconds. }
\centering
\begin{tabular}{|c|c|c|c|c|c|c|c|c|c|c|}
\hline
  $\lambda_i$ & $k$ & mass & $\Theta(k)$ & $\Phi(k)$&$\Phi(k)/(N+\sqrt{2N}) $ &$R^2$ \\ \hline
 $    10$ & $  200$ & $  2.51e+9$ & $   2.4959$ & $ 9563.2$ &$8.00$ & $  0.70 $\\ \hline
 $     0.25$ & $  200$ & $  3.18e+9$ & $   44.376$ & $  1910.6$ & $1.60$ &$  0.93$\\ \hline
 $    0.001$ & $   200$ & $  3.13e+9$ & $    21657$ & $  1485.4$ &$1.24 $ & $   0.42$ \\ \hline
 \end{tabular}
\label{tab6}
\end{table}


\section{Conclusions}\label{conclusion}
We have presented MATLAB software for $3$D inversion of gravity data using an equidistance stabilization term based on a graph theory argument that was  developed by \cite{Bijani:15}. The subsurface homogeneous body is  approximated by a set of point masses that provide a skeleton of  a subsurface structure. The  point masses are associated  with a full graph and  Kruskal's algorithm is used to find the minimum spanning tree of the graph.    The equidistance stabilization term  restricts the spatial distribution of the point masses and suggests a homogeneous spatial distribution of point masses in the subsurface. The global objective function is minimized using a  genetic algorithm using crossover, mutation and random population initialization, with a priori constraints on the parameters imposed at all stages of the population evolution.  A module for generating a synthetic geometry and gravity data set is also provided. The software is user-friendly and can be modified  to use for practically acquired data sets and simulations of synthetic data.  It is open source software and available at  \cite{VANRK:18}. 

The software was illustrated for a  physically realistic test problem with Gaussian noise added to the gravity measurements. The objective function includes a regularization parameter which balances the relative importance of the data misfit and the equidistance stabilization during the optimization. It was demonstrated that a suitable choice of regularization parameter is one for which (i) the predicted data are close to the observed data relative to the noise level and (ii) the equidistance function converges almost monotonically to zero with increasing numbers of iterations.  Thus it is sufficient to carry out the optimization for relatively few choices of $\lambda$, particularly when similar data sets have been previously analyzed and an acceptable range for the  regularization parameter  has been found. To assist with identification of $\lambda$ a new statistical approach based on linear regression analysis   has been illustrated and assists with identification of $\lambda$ when no prior data sets have been analyzed. 

The methodology was illustrated for  gravity data from the Mobrun ore body. The maximum extensions of the body in the east and north directions were found to be approximately $350$~m and $200$~m, respectively, and are in good agreement with results from previous investigations and from drill hole information.

\section*{Acknowledgements}
 R.A. Renaut acknowledges the support of NSF grant DMS $1418377:$ ``Novel Regularization for Joint Inversion of Nonlinear Problems". 
\begin{appendices}
\section{Genetic Algorithm Parameters}\label{GAparams}
\begin{table}
\begin{center}
\begin{tabular}{|c|c|c|c|c|c}   \hline
Population Size&  noq \\ \hline
Max Generations &  $K_{\mathrm{max}}$ \\\hline
Cross Over Percentage&  CP \\ \hline
Extra Range Factor for Crossover&  Errf \\ \hline
Mutant Percentage&  MP \\ \hline
Mutation Rate & $\mu$ \\ \hline
Selection Pressure & $\beta$ \\ \hline
Number of Point Masses & M \\ \hline
Minimum total mass &  $m_{t_{\mathrm{min}}}$ \\ \hline
Minimum in East Direction & $x_{{\mathrm{max}}}$\\ \hline
Minimum in North Direction &  $y_{{\mathrm{max}}}$\\ \hline
Minimum in Depth Direction &  $z_{{\mathrm{max}}}$\\ \hline
Maximum total mass &  $m_{t_{\mathrm{max}}} $\\ \hline
Maximum in East Direction &  $x_{{\mathrm{max}}}$ \\ \hline
Maximum in North Direction & $y_{{\mathrm{max}}}$\\ \hline
Maximum in Depth Direction &$ z_{{\mathrm{max}}}$ \\ \hline
\end{tabular}
\caption{Input Parameters used for the Genetic Algorithm. \label{Tab:OptimParam}}
\end{center}
\end{table}
\end{appendices}

\end{document}